\documentclass[a4paper,11pt]{article}
\usepackage{jheppub} % for details on the use of the package, please see the JINST-author-manual
\usepackage{lineno}
\usepackage{graphicx}
\usepackage{subcaption}
\usepackage{dcolumn}
\usepackage{bm}
\usepackage{tabularx}
\usepackage{amsmath}
\usepackage{xcolor}

\usepackage{amssymb}
\usepackage{enumitem}
\hypersetup{colorlinks = true,
            linkcolor = blue,
            anchorcolor =red,
            citecolor = blue,
            pdfauthor=author}
\usepackage[capitalise]{cleveref}

\title{\boldmath Diffusion Models as Stochastic Quantization in Lattice Field Theory}

\author[a,b,c]{L. Wang}
\affiliation[a]{Interdisciplinary Theoretical and Mathematical Sciences Program (iTHEMS), RIKEN, Wako, Saitama 351-0198, Japan}
\affiliation[b]{Shanghai Research Center for Theoretical Nuclear Physics, National Natural Science Foundation of China and Fudan University, Shanghai 200438, China}
\emailAdd{lingxiao.wang@riken.jp}
\affiliation[c]{Frankfurt Institute for Advanced Studies, Ruth Moufang Strasse 1, D-60438, Frankfurt am Main, Germany}

\author[d,e]{, G. Aarts}
\emailAdd{g.aarts@swansea.ac.uk}
\affiliation[d]{Department of Physics, Swansea University, SA2 8PP, Swansea, United Kingdom}
\affiliation[e]{European Centre for Theoretical Studies in Nuclear Physics and Related Areas (ECT*) \& Fondazione
Bruno Kessler Strada delle Tabarelle 286, 38123 Villazzano (TN), Italy}

\author[f,c,1]{and K. Zhou \note{Corresponding author.}}
\affiliation[f]{School of Science and Engineering, The Chinese University of Hong Kong, Shenzhen (CUHK-Shenzhen), Guangdong, 518172, China}
\emailAdd{zhoukai@cuhk.edu.cn}

\abstract{In this work, we establish a direct connection between generative diffusion models (DMs) and stochastic quantization (SQ). The DM is realized by approximating the reversal of a stochastic process dictated by the Langevin equation, generating samples from a prior distribution to effectively mimic the target distribution. Using numerical simulations, we demonstrate that the DM can serve as a global sampler for generating quantum lattice field configurations in two-dimensional $\phi^4$ theory. We demonstrate that DMs can notably reduce autocorrelation times in the Markov chain, especially in the critical region where standard Markov Chain Monte-Carlo (MCMC) algorithms experience critical slowing down. The findings can potentially inspire further advancements in lattice field theory simulations, in particular in cases where it is expensive to generate large ensembles.}


\begin{document}
\maketitle
\flushbottom

\section{Introduction}

To obtain physical observables in lattice field theory, it is essential to approximate the path integral as a sum over field configurations. Monte Carlo methods rely on random sampling from the probability distribution, determined by the action of the system~\cite{Knechtli:2017sna}. These methods can have high computational costs, particularly in the vicinity of a critical point in parameter space~\cite{Wolff:1989wq}. Generative models, as a class of machine learning (ML) algorithms, can be trained to generate new data that follow the same distribution as a given data set, representing the underlying target distribution~\cite{tomczak:2022deep}. 
An important potential application of generative models in lattice QCD is therefore to improve the efficiency of Monte Carlo simulations~\cite{Boyda:2022nmh}. 

Depending on the way the likelihood is estimated, roughly two main categories of generative models are being developed. In implicit maximum likelihood estimation (MLE), Generative Adversarial Networks (GANs) can learn to generate new configurations via a min-max game using training on a dataset prepared using, e.g., a standard lattice simulation. 
A well-trained GAN can subsequently generate new samples, which should follow the same statistical distribution as the original dataset, but with reduced computational cost. Additionally, GANs can be used to improve the interpretability of lattice QCD simulations by generating visualizations of the system's behavior~\cite{Zhou:2018ill,Pawlowski:2018qxs}. 
In explicit MLE, flow-based models have recently been proposed to improve the efficiency of lattice simulations~\cite{ Albergo:2019eim,  Kanwar:2020xzo, Nicoli:2020njz,Albergo:2021bna, Albergo:2021vyo, DelDebbio:2021qwf, Hackett:2021idh, Albergo:2022qfi, Bacchio:2022vje, Caselle:2022acb, Chen:2022ytr, Gerdes:2022eve, Albandea:2023wgd, Singha:2023cql}. Flow-based models are able to directly approach the underlying distribution of the quantum field without preparing training data. Although the method may encounter ``mode-collapse" and scalability problems~\cite{Abbott:2022zsh,Nicoli:2023qsl}, it allows for the generation of independent new configurations that can be used in MC simulations, again reducing the computational cost. The current progress in applying generative models also includes designing novel neural network architectures for specific field theories, e.g., autoregressive networks~\cite{Wang:2020hji,Luo:2023opo} and gauge-equivariant neural networks~\cite{Favoni:2020reg, Kanwar:2020xzo, Aronsson:2023rli, Abbott:2022zhs, Gerdes:2022eve,Luo:2023opo}. These approaches suggest that generative models have the potential to advance the capabilities of lattice QCD simulations. A comprehensive review can be found in Refs.~\cite{Zhou:2023pti, He:2023zin}.

Recently, the ML community has developed a new class of deep generative models known as Diffusion Models (DMs)~\cite{yang:2022diffusion}. Impressive success has been obtained in generating high-quality images via stochastic denoising processes, as showcased in prominent applications~\cite{Croitoru:2023dm} such as DALL-E 2~\cite{2022arXiv220406125R} and Stable Diffusion~\cite{Rombach_2022_CVPR}. As a prevalent class of generative models encoded in a stochastic process, DMs are grounded in Markov chains, and their implementation involves the utilization of variational inference methods. For an adoption in high-energy physics experiments, see, e.g., Refs.~\cite{Mikuni:2022xry,Mikuni:2023dvk}.

In numerical lattice field theory, there is an approach alternative to Monte Carlo methods to generate field configurations which relies on solving a stochastic differential equation, i.e., a Langevin equation. Its origin is stochastic quantization (SQ) as proposed by Parisi and Wu~\cite{Parisi:1980ys}. The basic idea of SQ is to view the quantum system as the thermal equilibrium limit of a hypothetical stochastic process with respect to a fictitious time. For gauge fields, it does not require gauge fixing~\cite{Parisi:1980ys}. Comprehensive reviews include Refs.~\cite{Damgaard:1987rr,Namiki:1993fd}.
For theories with a sign or complex weight problem, such as Quantum Chromodynamics at nonzero baryon density or theories in real (rather than Euclidean) time, the Langevin process can be extended to complex Langevin dynamics~\cite{Parisi:1983mgm}, which evades the sign problem in certain cases~\cite{Berges:2006xc,Aarts:2008rr,Aarts:2008wh,Seiler:2012wz,Sexty:2013ica}, see also the reviews~\cite{Aarts:2015tyj,Attanasio:2020spv,Berger:2019odf,Nagata:2021ugx}. Although the method has been put on firmer theoretical grounds~\cite{Aarts:2009uq,Aarts:2011ax,Nagata:2016vkn,Aarts:2017vrv,Scherzer:2018hid},
convergence and boundary condition issues continue to hinder obtaining a fully satisfying approach. Recent work to address this includes 
Refs.~\cite{WesthHansen:2022iqd,Alvestad:2021hsi,Alvestad:2022abf,Lampl:2023xpb}.

In this work, we first point out the correspondence between generative DMs from the ML community and SQ in quantum field theory. Then, we explore using DMs to generate lattice field configurations. Specifically, we demonstrate for a two-dimensional $\phi^4$ lattice field theory that, combined with a hybrid Monte Carlo (HMC) algorithm, DMs can serve as an efficient global sampler along a Markov chain, leading to shorter autocorrelation times. We provide self-consistency checks and numerical evidence that this approach is %statistically exact and 
able to capture the SQ perspective-induced stochastic dynamics of quantum field theories.

The paper is organized as follows. In the subsequent section, we provide a brief overview of the concept of SQ. In Sec.~\ref{sec:dm}, we present details of DMs. It is argued that the denoising process of DMs can explicitly serve as a SQ procedure. In Sec.~\ref{sec:phi4}, we explore the application of DMs to a two-dimensional scalar $\phi^4$ field theory.  Additionally, we demonstrate that DMs can accurately estimate the physical action in an unsupervised manner using the probabilistic flow approach. In Sec.~\ref{sec:sum}, we summarize the correspondence between DMs and SQ, and outline how this novel method can motivate further research, in particular for theories where it is expensive to sample from the prior distribution using standard approaches. App.~\ref{app:unet} contains information on the U-Net architecture employed, while Appendix \ref{app:she} contains a brief discussion on the Skilling-Hutchinson estimator.
A study of the dependence of observables and the acceptance rate on various system and DM parameters can be found in Appendix \ref{app:sta} and \ref{app:acc} respectively.

%%%%%%% SECTION %%%%%%%%%%%%%%%%%
\section{Stochastic Quantization}
\label{sec:sq}

In Euclidean quantum field theory, as an alternative quantization scheme, one may invoke stochastic quantization (SQ)~\cite{Parisi:1980ys}. We only consider real actions here, for which SQ is well understood and convergence is guaranteed \cite{Damgaard:1987rr}. Starting from a generic Euclidean path integral with Euclidean action $S_E$,
\begin{equation}
    Z = \int D\phi \, e^{-S_E},
\end{equation}
SQ introduces a \textit{fictitious} time, $\tau$, for the field, $\phi$, whose evolution is described by Langevin dynamics,
\begin{equation}
    \frac{\partial \phi(x,\tau)}{\partial \tau} = - \frac{\delta S_E[\phi]}{\delta \phi(x,\tau)} + \eta(x,\tau),
    \label{eq:sq}
\end{equation}
where the noise term routinely satisfies,
\begin{equation}
    \langle \eta(x,\tau) \rangle = 0,\qquad \langle \eta(x,\tau)\eta(x',\tau') \rangle = 2\alpha\delta(x-x')\delta(\tau - \tau'),
\end{equation}
with $\alpha$ being the diffusion constant. Denoting the distribution as
\begin{equation}
    p(\phi) = \frac{e^{-S_E(\phi)}}{Z},
\end{equation}
the drift term in the Langevin equation can be written as
\begin{equation}
\label{eq:K}
    K(\phi)  \equiv  - \frac{\delta S_E}{\delta \phi(x)} = \frac{\delta \log p(\phi)}{\delta \phi(x)}.
\end{equation}
In the long-time limit ($\tau\rightarrow \infty$) the system reaches an equilibrium state of the dynamics~\cite{Namiki:1993fd} if the drift term has a damping effect, which is best analysed using the Fokker-Planck equation (see below). In addition, introducing the stochasticity is only one strategy for the quantization. The \textit{fictitious} time itself can serve as a clue to develop microcanonical quantization, which describes a deterministic evolution with an ordinary differential equation~\cite{Callaway:1982eb}. We will revisit this idea and its connection with DMs in Sec.~\ref{subsec:continious}.

\subsection{Fokker-Planck Equation and Observables}

The Fokker-Planck equation~\cite{risken1996fokker} describes the time evolution of the probability distribution of field configurations, subject to a stochastic process, such as the Langevin equation. Also known as the Kolmogorov forward equation, in the context of SQ it describes the time evolution of the probability density function $P[\phi,\tau]$ for field configurations $\phi$ at time $\tau$ \cite{Damgaard:1987rr,Namiki:1992wf}, as follows,
\begin{equation}
 \frac{\partial P[\phi, \tau]}{\partial \tau} = \int d^nx \left\{
 \frac{\delta}{\delta \phi}
 \left(\alpha\frac{\delta }{\delta \phi} + \frac{\delta S_E}{\delta \phi} \right)
 \right\} P[\phi, \tau].
 \label{eq:f-p}
\end{equation}
Here $n$ is the dimension of the Euclidean spacetime on which the field theory is formulated.  Given this probability distribution function, the expectation value of an observable $\mathcal{O}[\phi]$ is then given by \cite{Damgaard:1987rr}
\begin{equation}
\langle \mathcal{O}[\phi] \rangle_\tau = \int D\phi\,\mathcal{O}[\phi] P[\phi, \tau],
\end{equation}
where the left-hand side is understood as an average over many stochastic processes and on the right-hand side the integral is taken over all field configurations $\phi$. 

In the long-time limit, the probability distribution $P[\phi,t]$ {approaches} an equilibrium distribution, which is the stationary solution of the Fokker-Planck equation (\ref{eq:f-p}), 
\begin{equation}
P_{\text{eq}}[\phi] \propto e^{-\frac{1}{\alpha}S_E[\phi]}.
\label{eq:equ}
\end{equation}
If $\alpha$ is set (formally) as $\hbar$ and the equilibrium distribution is normalised, expectation values of observables are indeed observables in a quantum theory, with
\begin{equation}
\langle \mathcal{O}[\phi] \rangle_{\tau\to\infty} = \frac{\int D\phi \, \mathcal{O}(\phi) e^{-\frac{1}{\hbar}S_E(\phi)}}{\int D\phi \, e^{-\frac{1}{\hbar}S_E(\phi)}}
= \langle \mathcal{O}[\phi] \rangle_{\text{quantum}},
\end{equation}
demonstrating that a stochastic process according to Eq.~(\ref{eq:sq}) can be used to quantize a system.
% Note that convergence to the stationary solution (\ref{eq:equ}) follows from properties of the Fokker-Planck Hamiltonian associated to the Fokker-Planck equation (\ref{eq:f-p}) \cite{Damgaard:1987rr}. 

{
Note that convergence to the stationary solution (\ref{eq:equ}) follows from properties of the Fokker-Planck Hamiltonian associated to the Fokker-Planck equation (\ref{eq:f-p}) \cite{Damgaard:1987rr}. In short, writing the time-dependent solution $P[\phi,t]$ as 
\begin{equation}
\label{eq:PQ}
P[\phi,t]=e^{-\frac{1}{2\alpha}S_E[\phi]} Q[\phi,t],
\end{equation}
it is easy to show that $Q[\phi,t]$ satisfies the equation
\begin{equation}
\label{eq:Q}
\frac{\partial Q[\phi, \tau]}{\partial \tau}  = -H_{\rm FP}Q[\phi, \tau],
\end{equation}
where the Fokker-Planck Hamiltonian reads
\begin{equation}
H_{\rm FP} = \int d^nx\, L^\dagger L,
\end{equation}
with
\begin{equation}
L = \sqrt{\alpha}\frac{\delta}{\delta \phi} + \frac{1}{2\sqrt{\alpha}}\frac{\delta S_E}{\delta\phi}, 
\qquad\qquad
L^\dagger = -\sqrt{\alpha}\frac{\delta}{\delta \phi} + \frac{1}{2\sqrt{\alpha}}\frac{\delta S_E}{\delta
\phi}. 
\end{equation}
The Fokker-Planck Hamiltonian is hence bounded from below by zero. Provided its spectrum is gapped, time-dependent components in Eq.~(\ref{eq:Q}) will decay exponentially and only the solution $H_{\rm FP}Q[\phi]=0$ survives. This solution is given by
\begin{equation}
LQ[\phi]=0 \qquad \Rightarrow \qquad Q[\phi] \propto 
e^{-\frac{1}{2\alpha}S_E[\phi]},
\end{equation}
which, when combined with Eq.~(\ref{eq:PQ}),
results again in the equilibrium solution (\ref{eq:equ}), completing the standard proof of convergence.
}

\subsection{Langevin Simulations}
\label{subsec:sim}

In most cases, the Fokker-Planck equation cannot be solved analytically. Instead, one solves the stochastic process numerically by discretizing the Langevin equation Eq.~\eqref{eq:sq}. 
Using It\^o calculus, the simplest discretized Langevin equation, using time step $\Delta\tau$, is
\begin{equation}
    \phi(x,\tau_{n+1}) = \phi(x,\tau_n) + K[\phi(x,\tau_n)]\Delta\tau + \sqrt{\Delta\tau}\eta(x,\tau_n),
    \label{eq:dis-sq}
\end{equation}
with $\langle \eta(x,\tau_n)\eta(x',\tau_{n'}) \rangle = 2\alpha\delta(x-x')\delta_{nn'}$. 

\begin{figure}[thbp!]
    \centering
    \includegraphics[width = 0.8\textwidth]{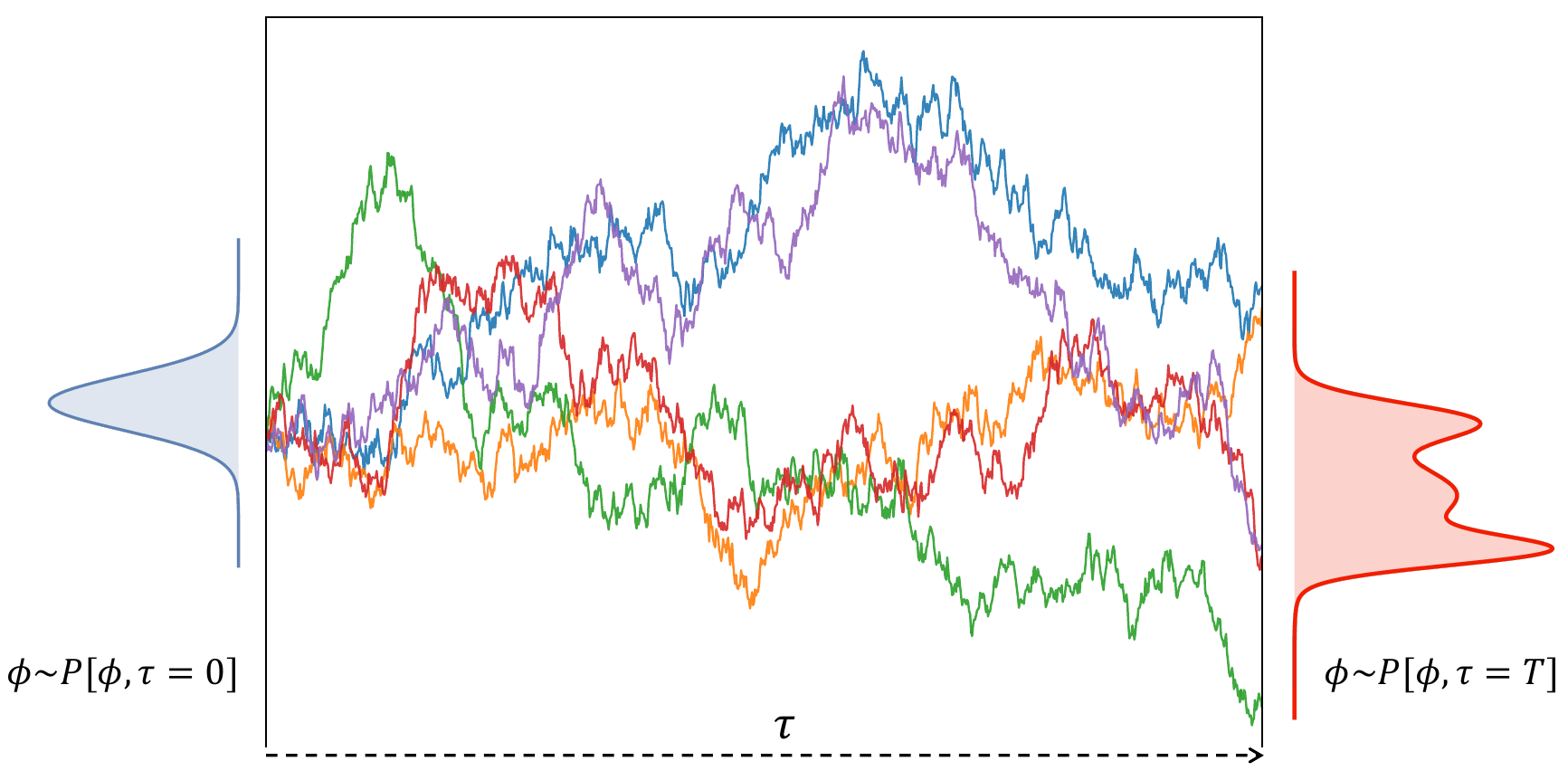}
    \caption{Sketch of a Langevin simulation. Each line represents a different trajectory, evolving from an initial state at $\tau=0$ to one at a later time, $T$. The initial configurations are sampled from a prior but naive distribution, $\phi\sim P[\phi,\tau=0]$. After the simulation, the final configurations follow the equilibrated distribution, $\phi\sim P[\phi,\tau=T]\sim P_{\rm target}[\phi]$.}
    \label{fig:simu}
\end{figure}

A typical calculation includes, firstly, an initial condition to prepare field configurations $\{\phi(x, \tau=0)\}$; secondly, the thermalization procedure, i.e, evolve the system to reach equilibrium by iterating through a sufficient number of time steps with an appropriate time step, $\Delta \tau$; thirdly, after thermalization, record the values of the physical observable $\mathcal{O}[\phi]$ for each configuration in an ensemble, e.g.\ after a time $T$. This is repeated for numerous configurations to improve the statistical accuracy of the ensemble average. 
Fig.~\ref{fig:simu} demonstrates one such typical Langevin process for a one-dimensional case. 
The target distribution is encoded via the drift, c.f.\ Eq.~(\ref{eq:K}).

%%%%%%%%%%%%%%%%%%%%%%%%%%%%%%%%%%%%%%%%%%%%%%%%%%%
%%%%%%%%%%%%%%%%%%%%%%%%%%%%%%%%%%%%%%%%%%%%%%%%%%%

{This procedure provides an estimate of the quantum expectation value for the observables $\mathcal{O}[\phi]$. Since the simple discretization scheme above results in finite-stepsize errors linear in $\Delta\tau$, one has to repeat the analysis to extrapolate to vanishing stepsize. The Langevin simulation provides an alternative to Monte Carlo approaches and has been used in simulations of lattice QCD~\cite{Batrouni:1985jn}. A characteristic is that no accept/reject step is used, which makes the algorithm not exact (as demonstrated by e.g.,\ the finite-stepsize corrections). This can be remedied by introducing such an accept/reject step, which results in more sophisticated algorithms that outperform the basic Langevin algorithm and are hence favoured (such as Hybrid Monte Carlo). The absence of an accept/reject step is turned into a feature for theories with a complex-valued Boltzmann weight, for which importance sampling is not possible. The application of SQ to these theories is commonly referred to as complex Langevin dynamics~\cite{Parisi:1983mgm}, which is still an active area of research~\cite{Berges:2006xc,Aarts:2008rr,Aarts:2008wh,Seiler:2012wz,Sexty:2013ica,Aarts:2015tyj,Attanasio:2020spv,Berger:2019odf,Nagata:2021ugx,Aarts:2009uq,Aarts:2011ax,Nagata:2016vkn,Aarts:2017vrv,Scherzer:2018hid,WesthHansen:2022iqd,Alvestad:2021hsi,Alvestad:2022abf,Lampl:2023xpb},
but not further explored here.}

%%%%%%% SECTION %%%%%%%%%%%%%%%%%

\section{Diffusion Models}
\label{sec:dm}

Turning now to deep learning, DMs denote a class of deep generative algorithms that simulate a stochastic process to generate samples following a desired target distribution~\cite{sohl-dickstein:2015deep}. The fundamental concepts underlying DMs involve modeling the data distribution through a continuous-time diffusion process or a discrete-time Markov chain, wherein noise is incrementally added to or removed from the data. This generative framework has gained prominence in recent works, such as score-based models (SBMs)~\cite{NEURIPS2019_3001ef25,song2021scorebased} and denoising diffusion probabilistic models (DDPMs) \cite{ho:2020denoising}; for a comprehensive review, see Ref.~\cite{yang:2022diffusion}. In this section, we will first introduce the denoising perspective of DMs and then point out the connections of DMs to SQ.

\subsection{Denoising Model}

In the context of DMs, the denoising model can be perceived as a mapping that reconstructs the original data from its noisy variant, which is attained through the application of a diffusion process. Similar to the Langevin dynamics introduced above, the first process (also called forward process) gradually introduces noise to the data, effectively ``smoothing'' the underlying (but typically unknown) probability distribution of data. Then the denoising model aims to learn its \textit{inverse process}, i.e., denoising the data by eliminating the added noise. Upon training, the denoising model can be employed to generate samples following the data distribution by executing the reverse diffusion process only. This sampling process commences with random noise and iteratively applies the trained denoising model to ``clean'' it until convergence, resulting in a sample representative of the target data distribution.

%%%%%%%%%%%%%%%%%%%%%%%%%%%%%%%%%%%%%%%%%%%%%%%%%%%%%%%%
\begin{figure}[tbp]
\begin{center}
\includegraphics[width=1.0\textwidth]{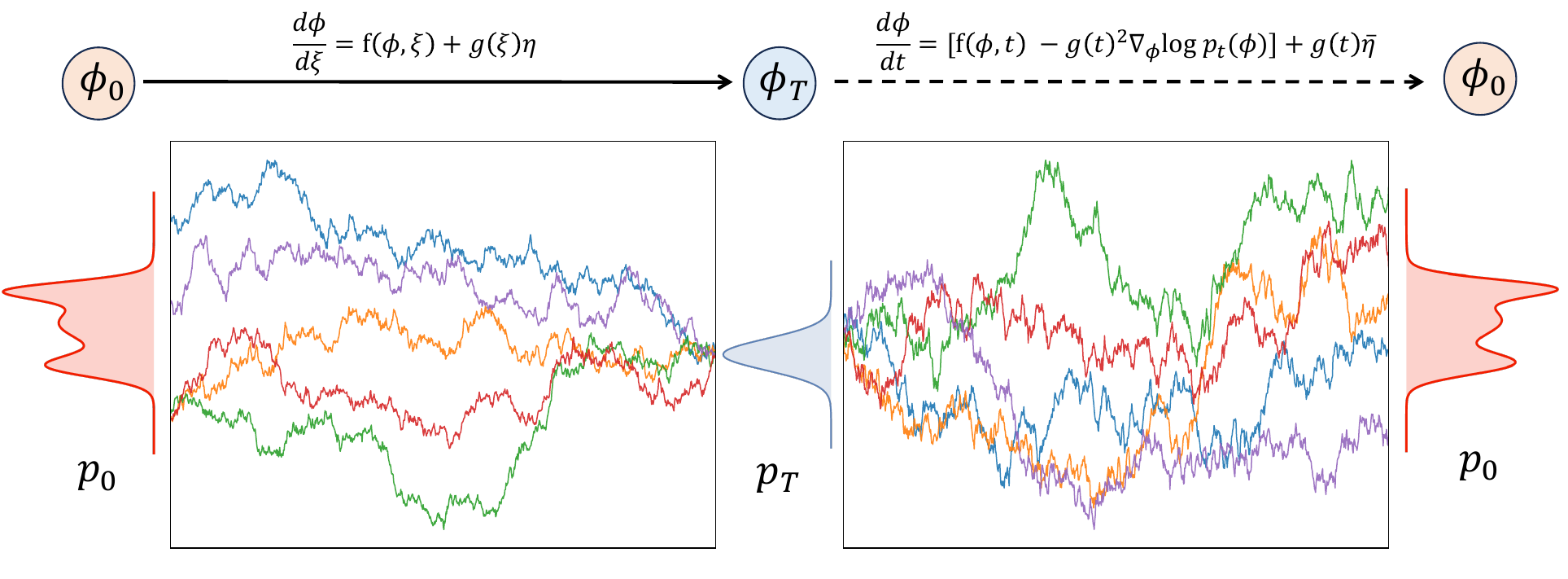}
\caption{A sketch of the forward diffusion process (left panel) and the reverse denoising process (right panel). The two stochastic processes are described by two stochastic differential equations. The target distribution is typically unknown but learnt from the training data.
}
\label{fig:dm}
\end{center}
\end{figure}
%%%%%%%%%%%%%%%%%%%%%%%%%%%%%%%%%%%%%%%%%%%%%%%%%%%%%%%%%%%

In the limit of continuous time, the forward process follows a stochastic differential equation (SDE),
\begin{equation}
\frac{d \phi}{d\xi } = f(\phi, \xi) + g(\xi) \eta(\xi),
\label{eq:sde}
\end{equation}
where $\xi\in[0,T]$ is the Langevin time, $\eta$ is the noise term, satisfying {$\langle \eta(\xi)\eta(\xi')\rangle=2\alpha \delta(\xi-\xi')$ with $\alpha \equiv 1/2$ as a convention in the machine learning community}, and $f(\phi, \xi)$ is the drift term. Note that the noise and drift term have the same size as the field $\phi$. Finally, the square of $g(\xi)$ is the time-dependent scalar \textit{diffusion coefficient}. 
The forward diffusion process can be modeled as the solution of such a generic SDE, provided the drift and diffusion coefficient are chosen appropriately.

As Fig.~\ref{fig:dm} demonstrates, the aim is to obtain a sample from the data distribution $p_0$ by starting from a sample of the prior distribution $p_T$ and reversing the above diffusion process. This is achieved by solving a reverse SDE, representing a diffusion process evolving backward in time, which reads~\cite{anderson:1982reversetime}
\begin{equation}
  \frac{d\phi}{dt} = \left[ f(\phi, t) - g^2(t)\nabla_{\phi}\log p_t(\phi) \right] 
  + g(t) \bar{\eta}(t),
  \label{eq:rsde}
\end{equation}
where the reverse time $t \equiv T-\xi$ and $\bar{\eta}$ is a noise term in the reverse time direction. Importantly, the drift term now contains two terms, including the gradient of the logarithm of $p_t(\phi)$, the probabilistic distribution at time $t$ in the forward diffusion process. This reverse SDE can be computed once we have specified the drift term and diffusion coefficient of the forward SDE, and determined the gradient of $\log p_t(\phi)$ for each $t\in[0, T]$.

The key to implementing a denoising model lies in accurately estimating the probability distribution, $p_t(\phi)$, under a well-designed forward diffusion process of Eq.~\eqref{eq:sde}. In score-based models~\cite{NEURIPS2019_3001ef25,song2021scorebased}, $\nabla_{\phi}\log p_t(\phi)$ is named as the \textit{score} of each sample. Routinely, one can introduce a perturbation kernel for each forward diffusion step\footnote{By 
adding to samples a sequence of Gaussian noise with scales, $\sigma_\text{min} = \sigma_1<\sigma_2<\cdots<\sigma_N = \sigma_\text{max}$, so the starting point is $p_1(\phi_1)\simeq p_0(\phi_0)$ when $\sigma_\text{min}$ is small enough, and the end point is $p_N(\phi)\simeq \mathcal{N}(\phi ; \mathbf{0}, \sigma_\text{max}^2 \mathbf{I})$ when $\sigma_\text{max}$ is large enough. Hereafter, we abbreviate $\phi(\xi)$ as $\phi_\xi$, or $\phi_i$ for the discretised case.} 
as $p_\xi(\phi_\xi | \phi_0)  \equiv \mathcal{N}(\phi_\xi ; \phi_0, \sigma_\xi^2 \mathbf{I})$, where $\phi_0\sim p_0 $. Thus, the \textit{score} can be computed from 
\begin{equation}
    p_{\xi}(\phi_\xi) = \int p_{\xi}(\phi_\xi|\phi_0) p_0(\phi_0) d\phi_0.    
\end{equation} 
A neural network $s_\theta(\phi,t)$ can be trained to approximate the \textit{score} conveniently, which is named as \textit{score matching}~\cite{Aapo:2005scb,Vincent:2011scb}. The parameters of the neural network are optimized with the objective,
\begin{equation}
    %\hat{\theta}=\arg\operatorname*{min}_{\theta} \sum_{i=1}^{N}\sigma_{i}^{2}\mathbb{E}_{p_0(\phi_0)}\mathbb{E}_{p_i(\phi_i\mid\phi_0)}\bigl[ \Vert\mathbf{s}_\theta(\phi_i,\xi)-\nabla_{\phi_i}\log p_i(\phi_i\mid\phi_0)\Vert_{2}^{2} \bigr].\label{eq:objective}
    \mathcal{L}_{\theta}= \sum_{i=1}^{N}\sigma_{i}^{2}\mathbb{E}_{p_0(\phi_0)}\mathbb{E}_{p_i(\phi_i\mid\phi_0)}
    \left[ \left\Vert s_\theta(\phi_i,\xi)-\nabla_{\phi_i}\log p_i(\phi_i|\phi_0)\right\Vert_{2}^{2}
    \right],
    \label{eq:objective}
\end{equation}
where the interval $T$ has been divided into $N$ segments with index $i$ (further details are given below). The optimal score-based model $s_{\hat{\theta}}(\phi,t)$ with $\hat{\theta}=\arg\operatorname*{min}_{\theta}\mathcal{L}_{\theta}$ should match the score at every time-step. Sampling turns to solving the reverse SDE shown in Eq.~\eqref{eq:rsde}, with the \textbf{score} replaced with $s_{\hat{\theta}}(\phi,t)$,
\begin{equation}
  \frac{d\phi}{dt} = \left[ f(\phi, t) - g^2(t)s_{\hat{\theta}}(\phi,t) \right] + g(t) \bar{\eta}(t).
\end{equation}

For a concrete and concise example, by putting the forward drift term $f(\phi,\xi)$ to zero and simplifying the diffusion coefficient, see, e.g., Ref.~\cite{song2021scorebased}, as $g(\xi) \equiv \sigma^{\xi}$, one gets a \textit{variation expanding} forward process whose transition kernel remains Gaussian,
\begin{equation}
    p_{\xi}(\phi_\xi | \phi_0) = \mathcal{N}\bigg(\phi_\xi; \phi_0, \frac{1}{2\log \sigma}(\sigma^{2\xi} - 1) \mathbf{I}\bigg).
    \label{eq:sbm}
\end{equation}

We note that both Eqs.~\eqref{eq:sde} and~\eqref{eq:rsde} are Langevin equations.\footnote{This approach has been utilized in Bayesian learning as stochastic gradient Langevin dynamics \cite{welling:2011bayesian}. In contrast to standard stochastic gradient-descent (SGD) algorithms, stochastic gradient Langevin dynamics introduces stochasticity into the parameter updates, thereby avoiding collapses into local minima.} In the context of the reverse SDE, given the gradients $\nabla_{\phi}\log p(\phi,t)$, it is convenient to generate samples from a prior distribution in {a stochastic evolution process, as introduced in Sec.~\ref{subsec:sim}.}

\subsection{Diffusion Model as Stochastic Quantization}

The reverse SDEs of DMs are mathematically related to the Langevin dynamics. To explore the connection of DMs to stochastic quantization in more detail, we choose the \textit{variance expanding} picture of the DMs. Its denoising process transfers $\phi_T$ back to $\phi_0$ with the same marginal densities as the forward process, but evolving according to
\begin{equation}
    \frac{d\phi}{dt} =  - g(t)^2 \nabla_\phi \log{p_t(\phi)} +  g(t) \bar{\eta}(t),
\end{equation}
where time $t$ flows backward from $T$ to 0. With the redefinition of time, $\tau\equiv T-t$ ($d\tau \equiv -dt$), and denoting $g_\tau=g(T-\tau)$, $q_{\tau}(\phi)=p_{T-\tau}(\phi)$ of the forward process, the SDE now reads
\begin{equation}
    \frac{d\phi}{d\tau} = g_\tau^2\nabla_\phi \log{q_{\tau}(\phi)} + g_\tau\bar{\eta}(\tau) \label{eq:dmle}\\ 
\end{equation}
and in discretised form
\begin{equation}    
    \phi(\tau_{n+1}) = \phi(\tau_n) +  g_{\tau_n}^2\nabla_\phi \log{q_{\tau_n}[\phi(\tau_n)]} \Delta\tau +  g_{\tau_n}\sqrt{\Delta\tau}\bar{\eta}(\tau_n),
\end{equation}
where the noise term $\langle\bar{\eta}(\tau_n)\rangle = 0, \langle\bar{\eta}(\tau_n)\bar{\eta}(\tau_{n'})\rangle = 2\bar\alpha\delta_{nn'}$, { with $\bar{\alpha} \equiv \hbar =1$ as a convention in quantum field theories}. Because the mean value of the noise term remains zero, the sign in front of the noise term is unrelated to the stochastic process. 
We note that $g^2(\tau)$ is referred to as a kernel, rescaling both the drift term and the variance of the noise term \cite{Damgaard:1987rr}.
Its effect can be absorbed by rescaling time with $g^2(\tau)$, or equivalently absorb it in the time step, $ g_\tau^2\Delta\tau$.

The derivation of the corresponding Fokker-Planck is straightforward and one finds
%More details of deriving the related Fokker-Planck equation can be found in App.~\ref{app:der-FP}. In essence, one can introduce a new noise scale, $\langle\bar{\eta}^2\rangle \equiv 2\bar{\alpha}$, as in the Langevin dynamics as described by Eq.\eqref{eq:sq}. Consequently, the Fokker-Planck equation for the reverse SDEs becomes,
\begin{equation}
    \frac{\partial p_{\tau}(\phi)}{\partial \tau} = 
    \int d^nx \left\{g^2_{\tau}  \frac{\delta}{\delta \phi}
    \left( \bar{\alpha}\frac{\delta }{\delta \phi} + \nabla_\phi S_\text{DM} \right)\right\} p_{\tau}(\phi),
    \label{eq:fp}
\end{equation}
where 
\begin{equation}
  \nabla_\phi S_\text{DM}\equiv -\nabla_\phi \log{q_{\tau}(\phi)}    
\end{equation}
is the score trained on the forward diffusion process, and $p_{\tau}(\phi)$ is the probability density of $\phi_{\tau}$ in the reverse diffusion process. One can immediately derive the following equilibrium condition, locally in time, 
\begin{equation}
    p_\text{eq}(\phi) \propto e^ {-S_\text{DM}/\bar{\alpha}},
    \label{eq:dmsq}
\end{equation}
where we recall that $\bar\alpha=1$ and that $S_\text{DM}$ is an effective action that will be learned in DMs. When the noise scales are on par with the quantum scale, that is, $O(\bar{\alpha})\sim O(\hbar)$, Eq.~\eqref{eq:dmsq} encompasses the quantum fluctuations exhibited in Eq.~\eqref{eq:equ}. In the $\tau\rightarrow T$ limit, $ p_{\tau=T}(\phi) \rightarrow P[\phi,T]$, where $P[\phi,T]$ is nothing but the time evolution of the PDF in SQ. This implies that the “equilibrium state” of SQ can be achieved by denoising from a naive distribution. Concurrently, sampling from a DM is equivalent to optimizing a stochastic trajectory to approach the “equilibrium state” in Eq.~\eqref{eq:equ} given a naive distribution from the SQ perspective.

However, reverting the noise configuration back to the physics configuration is a highly intricate process. In principle, one could solve the reverse SDE of Eq.~\eqref{eq:dmle} to depict this denoising process, while computing the ``time-dependent" drift term remains intractable. A particular type of deep neural network, known as the U-Net, is used here. The architecture of this network is discussed in more detail in the App.~\ref{app:unet}. The U-Net is tailored to parameterize the \textit{score function}, $\hat{\mathbf{s}}_\theta(\phi,\tau)$, for estimating this time-dependent drift term, $-\nabla_\phi \log{q_{\tau}(\phi)}$. It is designed to accept two inputs: time and a time-dependent configuration within a trajectory, while its output is the same size as the input (field configuration).

\subsection{{Building Effective Actions}}
\label{subsec:continious}

{To derive effective actions for quantum fields in DMs requires the \textit{probability flow} ODE formulation~\cite{Maoutsa:2020ode,song2021scorebased}, which provides a quantitative way to compute the negative log-likelihood of the field configuration.} Assuming a differentiable one-to-one mapping $\mathbf{h}$ that transforms a data sample $\phi_0 \sim p_0$ to one from a prior distribution $\mathbf{h}(\phi_0) = \phi_T \sim p_T$, the likelihood of $p_0(\phi_0)$ can be computed using the change-of-variable formula as
\begin{equation}
p_0(\phi_0) = p_T(\mathbf{h}(\phi_0)) |\operatorname{det}(J_\mathbf{h}(\phi_0))|,\label{eq:mapping}
\end{equation}
where $J_\mathbf{h}$ denotes the Jacobian of the mapping $\mathbf{h}$, and it is assumed that the prior distribution $p_T$ is easy to evaluate. ODE trajectories also define a one-to-one mapping from $\phi_0$ to $\phi_T$. In our case, {for quantum field theories,\footnote{The derivations can be found in Ref.~\cite{risken1996fokker} and Appendix D.1 of Ref.~\cite{song2021scorebased}. The difference for the factors in front of $g(t)^2$ is due to the convention of $\bar{\alpha}= 1$ in quantum field theories.} }
\begin{equation}
\frac{d\phi}{dt} = \left[f(\phi, t) - g(t)^2 \nabla_\phi \log p_t(\phi)\right].
\label{eq:ode}
\end{equation}
It aims to find the trajectory of the state variable $\phi$ that has the same marginal probability density $p_t(\phi)$ as the stochastic process described by the SDE, i.e., Eq.~\eqref{eq:sde}. An instantaneous change-of-variable formula can be derived to connect the probability of $p_0(\phi_0)$ and $p_T(\phi_T)$, as,{
\begin{equation}
p_0 (\phi_0) = \exp\left[\int_0^T dt\,\mathbf{\nabla}\cdot \tilde{f}(\phi_t, t) \right] p_T(\phi_T),\label{eq:jacobian}
\end{equation}
where the divergence is the trace of the Jacobian, with redefinition of $\tilde{f}(\phi_t, t)\equiv f(\phi, t) - g(t)^2 \nabla_\phi \log p_t(\phi)$}. The practical computation of the trace can be found in App.~\ref{app:she} by using the Skilling-Hutchinson estimator~\cite{grathwohl2018scalable}. 

From the perspective of probability current conservation, the stochastic process we derived in the previous section intrinsically connects with flow-based models, which have recently been widely explored in current lattice calculations~\cite{Albergo:2021vyo,Zhou:2023pti}. In a flow-based model, {one can construct a flow mapping with a neural network whose parameters are $\{\theta\}$, which represents a bijective transformation} between the physical distribution $p(\phi)$ and a prior distribution, e.g., 
$z\sim \mathcal{N}(\mathbf{0},\mathbf{I})$. Although it is common to build the flow mapping in discrete steps, {the transformation $\tilde{f}$ which follows,  
\begin{equation}
    \log p(\phi) = \log p(\mathbf{z}) - \log \left|\text{det}\frac{\partial \tilde{f}}{\partial \mathbf{z}}\right|, \label{eq:flow_mapping}
\end{equation}
}can also be designed in continuous steps.\footnote{In fact, if one enforces $\log p(\mathbf{z}) = \text{const.}$, the transformation becomes the \textit{trivializing} flow, see more details in Refs.~\cite{Luscher:2009eq,DelDebbio:2021qwf,Bacchio:2022vje,Albandea:2023wgd}.} It refers to the continuous normalizing flow~\cite{Gerdes:2022eve}, in which the parameterized mapping, {$\tilde{f}_{\hat{\theta}}$, is the solution to a neural ODE for time $t\in [0,T]$,
\begin{equation}
    \frac{d \phi(t)}{ dt} = \tilde{f}_\theta(\phi(t),t),\label{eq:flow_ode}
\end{equation}
with boundary conditions: $\phi(0) \equiv \mathbf{z}$, $\phi(T)\equiv \phi$. In which, a neural network with parameters, $\tilde{f}_\theta(\phi,t)$, can be introduced to represent the dynamical kernel. The probabilistic flow follows,

\begin{equation}
    \frac{d \log p(\phi(t))}{ dt} = - \left(\nabla_\phi\cdot \tilde{f}_\theta\right)(\phi(t),t)\label{eq:flow_action},
\end{equation}
with boundary $p(\phi(0)) \equiv p(\mathbf{z})$. Comparing the mapping and ODEs in~\cref{eq:mapping,eq:ode} with~\cref{eq:flow_mapping,eq:flow_ode}, one can immediately find that the flow mapping, $\tilde{f}_\theta(\phi(t),t)$, is equivalent to $\tilde{f}(\phi_t,t)$ in Eq.~\eqref{eq:jacobian} for DMs.}

\subsection{Effective Action in a Toy Model}

From Eqs.~\eqref{eq:dmle} and \eqref{eq:fp}, one can find a ``time-dependent" stochastic process. In  the variance-expanding DM, the forward process of Eq.~\eqref{eq:sde} features a predetermined time-dependent diffusion noise denoted as $\sigma^{T-\tau}$, which ensures field configurations will eventually be reduced to a noise configuration. The effective action and the corresponding denoising process can be understood from a field deformation perspective: \textit{the field is continuously deformed following a stochastic process from the noise distribution to the physical distribution}. 

%%%%%%%%%%%%%%%%%%%%%%%%%%%%%%%%%%%%%%%%%%%%%%%%%%%%%%%%%%%%%%%%%%%%%%%%%%%%%%%%%%%%%%%%%
\begin{figure}[hbtp!]
    \centering
    \includegraphics[width = 0.95\textwidth]{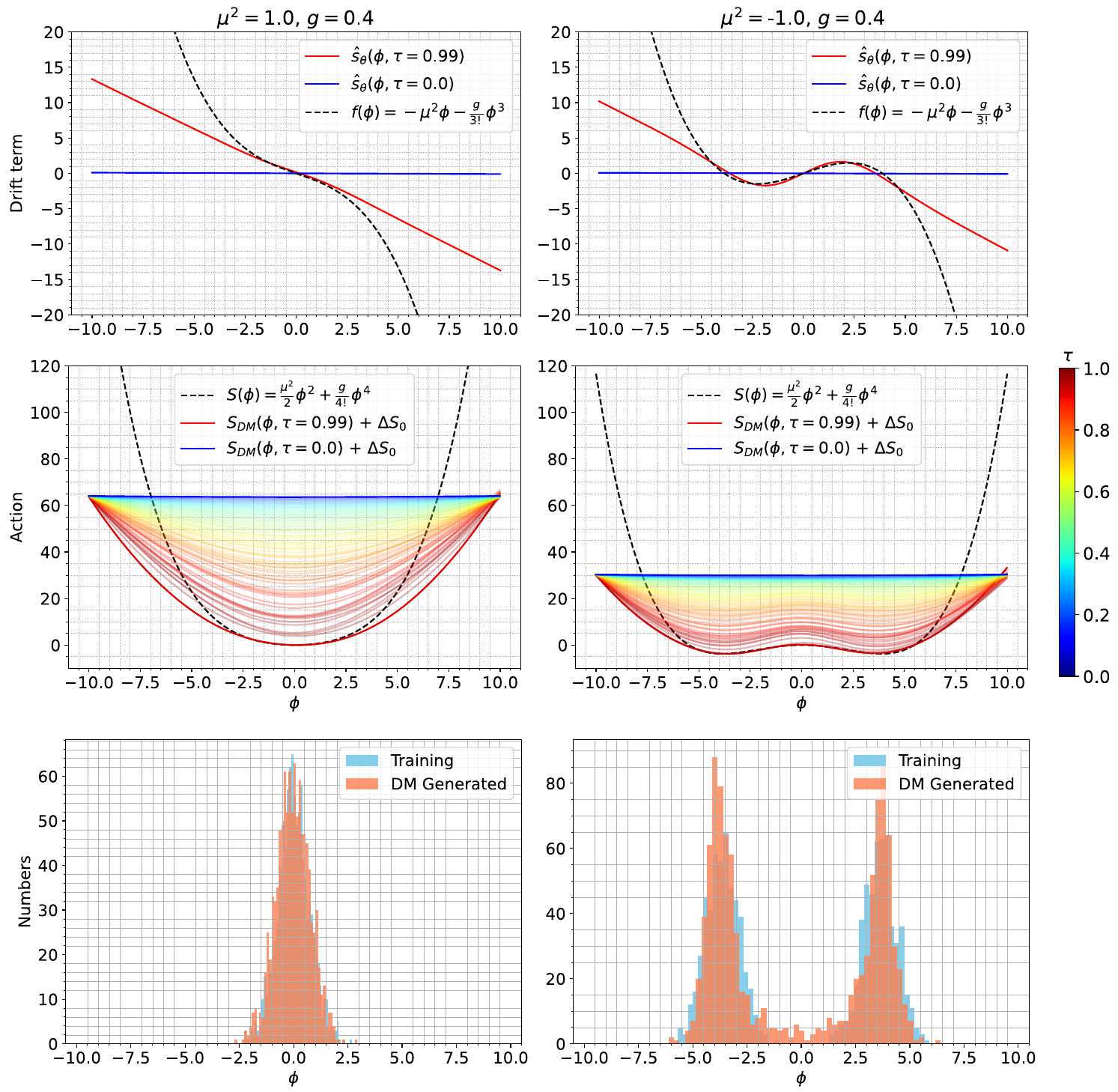}
    \caption{Drift terms (upper row) and effective actions (middle row) learned by the DM as a function of $\phi$ in both single-well (left column) and double-well (right column) actions, for various values of the time $\tau$ during the stochastic process. The action is shifted by a constant $\Delta S_0$. The dashed lines indicate the exact values. 1024 samples generated using the exact and the DM are shown in the bottom row. }
    \label{fig:learned}
\end{figure}
%%%%%%%%%%%%%%%%%%%%%%%%%%%%%%%%%%%%%%%%%%%%%%%%%%%%%%%%%%%%%%%%%%%%%%%%%%%%%%%%%%%%%%%%%

To demonstrate this in action, we introduce an over-simplified 0+0-dimensional field theory, i.e., a toy model which only has one degree of freedom,
\begin{equation}
    S(\phi) = \frac{\mu^2}{2}\phi^2 + \frac{g}{4!}\phi^4,
    \qquad
    f(\phi) = -\frac{\partial S(\phi)}{\partial\phi} = - \mu^2 \phi - \frac{g}{3!}\phi^3,
\end{equation}
where the physical action and drift term are determined by the parameters, $\mu^2$ and $g$. We prepared 5120 configurations as training data-sets in two set-ups: $\mu_1^2 = 1.0, g_1 = 0.4$ (single-well action), and $\mu_2^2 = -1.0, g_2 = 0.4$ (double-well action). A one-to-one neural network with time-embedding is implemented to represent the score function $\mathbf{s}_\theta(\phi,\tau)$.

After 500 epochs of training, the learned score function $\hat{\mathbf{s}}_\theta(\phi,\tau)$ is seen to approximate the drift term $f_\tau(\phi)$ in the upper row of Fig.~\ref{fig:learned}. 
Here the solid blue line represents the score function at the starting time ($\tau = 0$), 
the rainbow-colored lines indicate different $\tau > 0$ with a color-bar at rightest, while the solid red line indicates the score function at the end time ($\tau = 0.99$) in a backward process of the well-trained DM. 
The learned effective action $S_\text{DM}(\phi,\tau) = \int^{\phi}\hat{s}_\theta(\tilde{\phi},\tau) d\tilde{\phi} $ is shown in the middle row of Fig.~\ref{fig:learned}; it approximates the physical action as $\tau$ increases. We have shifted $S_\text{DM}$ with a constant $\Delta S_0$, which is the difference between $\min[S(\phi)]$ and $\min[S_\text{DM}(\phi,\tau)]$.  Generally, the learned drift terms and effective actions are accurate approximations in both the single and double-well cases, around $\phi\sim 0$. Samples generated with the trained DM are compared with samples from the underlying theory in the bottom row of Fig.~\ref{fig:learned}. 

The evolution trajectory of DM resembles an exact renormalizing group (RG) flow and its inversion. One intuitive understanding is that the gradually increasing noise scale (in the forward process of the DM) in coordinate space implicitly corresponds to a gradually decreasing momentum scale $\Lambda$ in an RG.\footnote{From Eq.~(\ref{eq:sbm}), one can sample to get the evolving field (for a given initial field $\phi_0$) at any time during the forward process as $\phi_{\tau}({\bf x})=\phi_{0}({\bf x}) + \sqrt{\frac{\sigma^{2\tau}-1}{2\log\sigma}}\epsilon({\bf x})$ with $\epsilon\sim\mathcal{N}({\bf 0},{\bf I})$ a Gaussian distributed random noise, which in momentum space is $\phi_{\tau}(p)=\phi_{0}(p)+\sqrt{\frac{\sigma^{2\tau}-1}{2\log\sigma}}\epsilon(p)$. Considering the natural distribution of momentum modes for the physical fields, the above evolution will perturb (smear) higher momentum modes faster because of the gradually increasing noise level. This resembles somehow the functional renormalization group (FRG) procedure to progressively integrate out the high momentum modes. Conversely, in the reverse denoising process, the low momentum IR physics will be first generated, and only at a later time the UV details are filled by the denoising model, as seen in Fig.~\ref{fig:sample}.} With $g_{\tau \rightarrow T} = \sigma^{T-\tau} \rightarrow 1$, $\Lambda \rightarrow \infty$, the full quantum theory is recovered. Although some works have discussed the running noise scales from a colored noise perspective in a Langevin process~\cite{Bern:1986xc,Pawlowski:2017rhn}, the well-trained DM provides an exact RG flow in the \textit{fictitious} time space,
\begin{equation}
    \frac{\partial S_\text{DM}(\phi,\tau)}{\partial \tau} = \frac{\partial}{\partial \tau} \int^{\phi}\hat{s}_\theta(\tilde{\phi},\tau) d\tilde{\phi}.
\end{equation}
Because the probability density is positive-definite, the equation has a fixed point when $\partial p_\tau(\phi)/\partial\tau = 0$.  The RG flow is the evolution trajectory of the system from the trivial distribution at $\tau = 0$ to the equilibrium state at $\tau = T$, which has been shown in the middle panel of Fig.~\ref{fig:learned}.

%%%%%%% SECTION %%%%%%%%%%%%%%%%%

\section{Diffusion Models for Lattice $\phi^4$ Scalar Fields}
\label{sec:phi4}

Lattice scalar field theories have been studied extensively, with applications spanning a wide range of topics. Scalar fields are frequently employed to showcase the efficacy of numerical algorithms, including with machine learning. Here we will implement a DM to generate configurations for a two-dimensional scalar field theory with quartic interaction, both in the broken and the symmetric phase. 

We consider a real scalar field in $d$ \textit{Euclidean} dimensions with the action, 
\begin{equation}
S_E = \int \text{d}^{d}x\left( \frac{1}{2}\sum_{\mu=1}^d\left(\partial_\mu\phi_0\right)^2 + \frac{1}{2} m_0^2\phi_0^2 + \frac{\lambda_0}{4!}\phi_0^4 \right),
\end{equation}
where the subscript specifies bare quantities, including mass $m_0$, coupling $\lambda_0$, and field $\phi_0(x)$. Discretising the derivative on a lattice with lattice spacing $a$ in all directions,   $\partial_\mu\phi_0(x)=[\phi_0(x+a\hat\mu)-\phi(x)]/a$, with $\hat{\mu}$ the unit vector along the $\mu-$direction, and redefining the field and parameters to yield dimensionless combinations, gives the lattice action (see e.g.\ Ref.~\cite{Smit:2002ug})
\begin{equation}
S_E = \sum_x \left[  -2\kappa \sum_{\mu = 1}^d\phi(x)\phi(x+\hat{\mu}) + (1-2\lambda)\phi^2(x) + \lambda\phi^4(x) \right],
\label{eq:phi4action}
\end{equation}
where $\kappa$ is the hopping parameter, related to the bare mass parameter via
\begin{equation}
     (am_0)^2 = \frac{1-2\lambda}{\kappa} - 2d,
\end{equation}
and $\lambda=a^{4-d}\kappa^2\lambda_0/6$ denotes the dimensionless coupling constant describing field interactions. Both parameters are positive. The field has been rescaled according to $a^{d/2-1}\phi_0=(2\kappa)^{1/2}\phi$.

In the case of $d\geq2$, for each coupling $\lambda$ the hopping parameter exhibits a critical value $\kappa_c(\lambda)$, at which a second-order phase transition occurs. This quantum phase transition corresponds to a spontaneous $\mathbb{Z}_2$ symmetry broken above the critical point in continuum limit~\cite{Akiyama:2021zhf}. When the mass term becomes negative, the broken phase emerges classically.  At the classical level, the critical value for the hopping parameter is given by
\begin{equation}
    \kappa_\text{c}^{\rm cl}(\lambda) = \frac{1-2\lambda}{2d},\label{eq:kc}
\end{equation}
which is determined by the vanishing of the mass term,  but quantum fluctuations will change the value \cite{Smit:2002ug}. As $\kappa$ decreases across the critical value, the system transitions from the broken phase to the symmetric phase.

\subsection{Configuration Generation and Model Setups}

To implement the DM, we first prepare field configurations of the scalar $\phi^4$-theory in $d=2$ dimensions on a $32 \times 32$ lattice, at fixed $\lambda = 0.022$, as defined above. To evaluate the performance of the model, we generated 5120 samples for training and 1024 samples for testing in the broken and symmetric phases at $\kappa = 0.5$ and $\kappa = 0.21$ respectively. The reason to choose these specific paramter values is that at $\lambda = 0.022$ the phase transition occurs at approximately $\kappa_c \simeq 0.239$, as derived from Eq.~\eqref{eq:kc}.

\textbf{Hybrid Monte Carlo set-up.} In order to prepare the dataset, we employ the Monte Carlo Markov Chain (MCMC) approach. The classical Metropolis-Hastings algorithm for local updating often suffers from long autocorrelation times, so we instead opt for the hybrid Monte Carlo (HMC) method. By incorporating classical Hamiltonian equations of motion~\cite{neal2011mcmc}, the HMC method improves the performance of the MCMC and allows larger steps to be taken while maintaining acceptable acceptance rates. In our setup, using a chosen set of action parameters, we prepare a 2-dimensional lattice of size $32 \times 32$. For each trajectory in the HMC, we initiate with a "100-trajectory burn-in sequence" for thermalization. Following this, after 64 additional trajectories, we select the outcome of the final trajectory update as a configuration. Each trajectory utilizes a well-established HMC integrator such as the leapfrog, with a certain stepsize $\Delta t$. The total length of each trajectory is determined by $\tau = n_\text{step} \times \Delta t$, with $n_\text{step}$ representing the number of times the integrator is applied within a single HMC trajectory (commonly set so that $\tau = 1$). Each configuration in the training dataset originates from one different Markov chain.

\textbf{Diffusion Model set-up.} We adopt the variance expanding DM as shown in Eq.~\eqref{eq:sbm} with $\sigma = 25$. The diffusion time is set as $T = 1$. 
In the training procedure the stepsize is not fixed, but we randomly sample the time points in the interval $[0,T]$, with the number of time points equal to the batch size in each batch optimization. The batch size is 64, and the maximum number of epochs for training is 250 to ensure convergence.  In the sampling procedure, we use the Euler-Maruyama approach to solve the reverse-time SDE with time stepsize $T/N = 0.002$. The U-Net architecture~\cite{Ronneberger:2015unet} is utilized to represent $\mathbf{s}_\theta$ in Eq.~\eqref{eq:objective}; more details can be found in App.~\ref{app:unet}. Unless otherwise indicated, we generated 1024 samples from the well-trained DM to demonstrate its performance.

{We deployed the HMC and DM algorithms on different devices. For the chosen parameter values ($\kappa = 0.21, \lambda = 0.022, 32\times 32$ lattice), the HMC generates samples on CPU (Intel Xeon Gold 6226R @ 2.90GHz) with 1.57 configuration/s. The DM generates samples on GPU (Nvidia GeForce RTX 3090 @ 1.40Ghz) with 97.52 configuration/s.}

\subsection{Broken Phase}

\begin{figure}[!htbp]
    \begin{minipage}{0.50\textwidth}
    \centering
    \includegraphics[width=\textwidth]{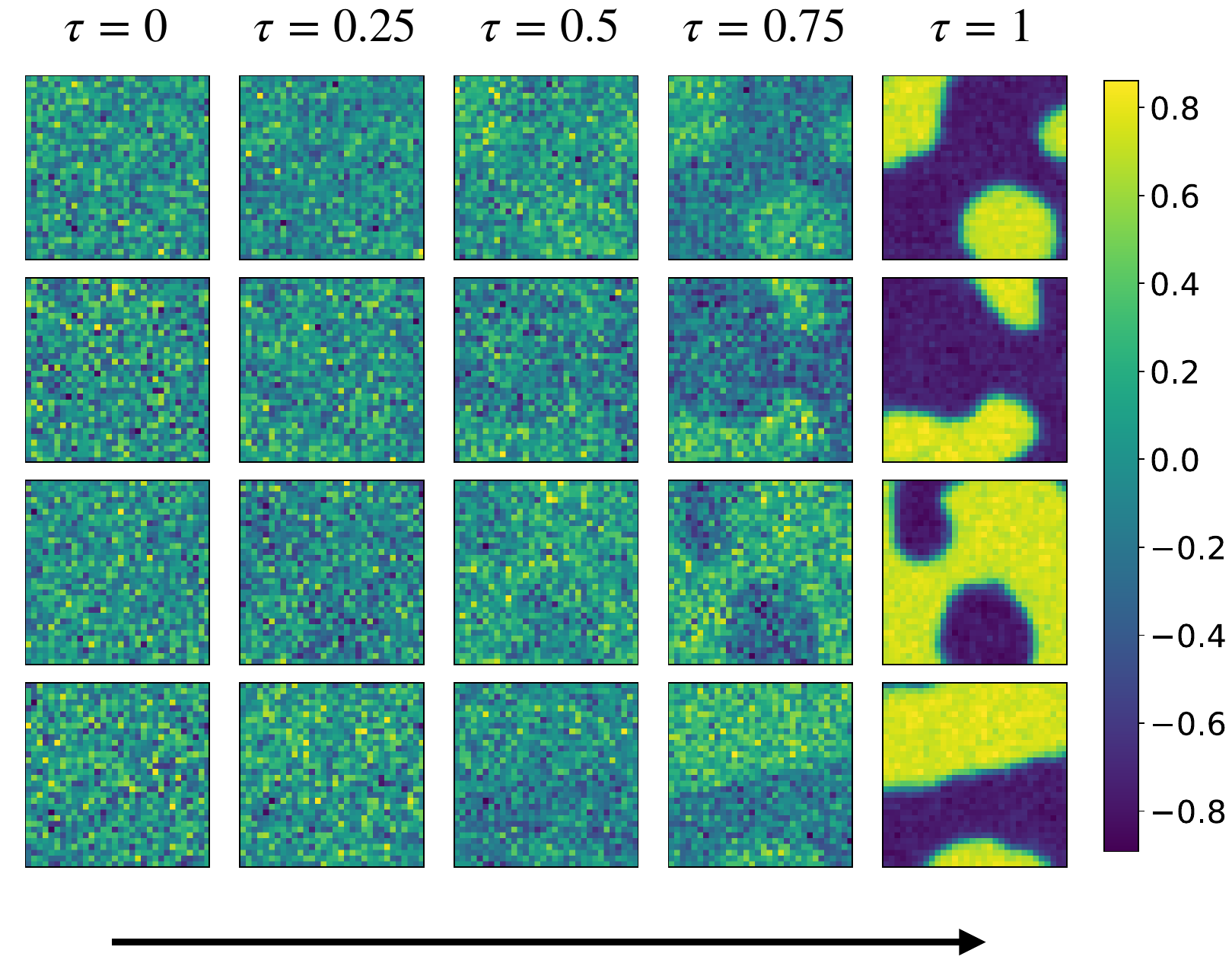}
    \caption{The denoising process for generating four independent configurations from a well-trained DM.}\label{fig:sample}
    \end{minipage}
\hfill
    \begin{minipage}{0.45\textwidth}
    \centering
    \includegraphics[width=\textwidth]{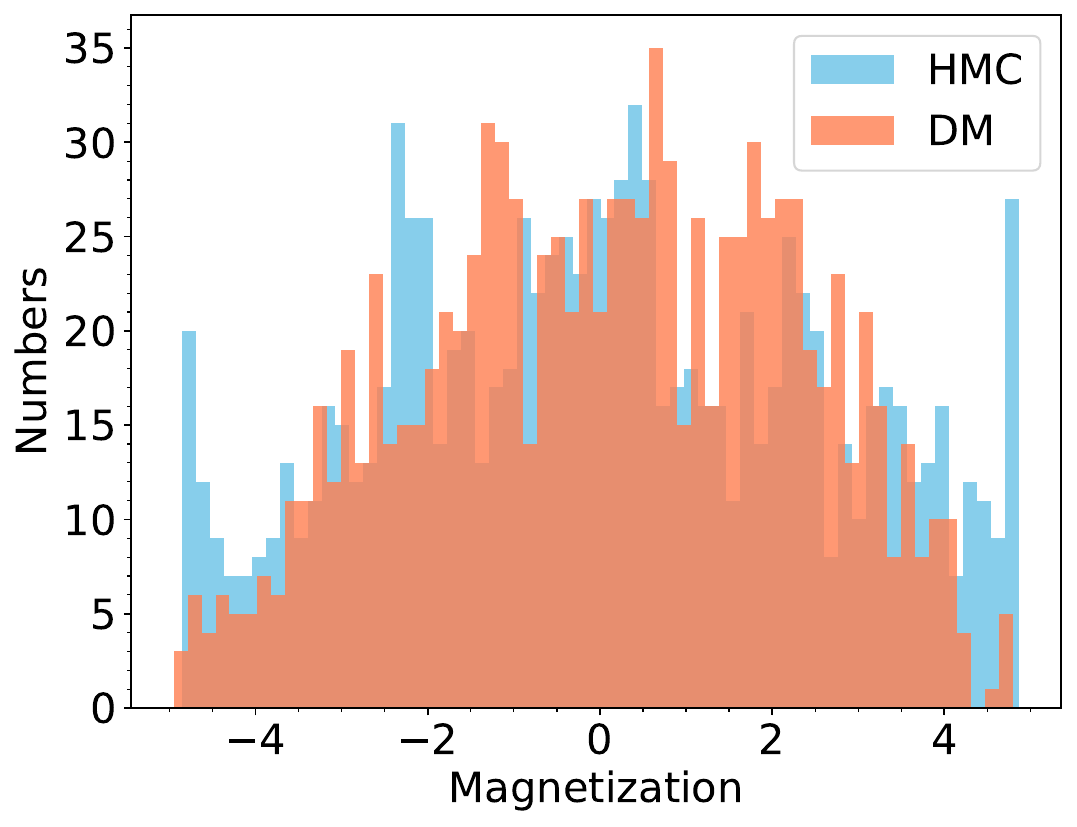}
    \caption{Comparison of the distribution of the magnetization with 1024 samples generated from the well-trained DM and using HMC.}\label{fig:brokenphase}
    \end{minipage}
\end{figure}

We start in the broken phase. Here one expects large clusters, in which field configurations fluctuate around a positive or negative average value, usually referred to as the vacuum expectation value, order parameter or magnetisation. We demonstrate that the clustering behavior of field configurations can be successfully captured by the well-trained DM. Fig.~\ref{fig:sample} visualizes the denoising process. Each row in the figure represents a different sample, and each column represents a different time point in the denoising process. The first column represents noise samples randomly drawn from the prior distribution, i.e., $p \approx \mathcal{N}\bigg(\phi; \mathbf{0}, \frac{1}{2\log \sigma}(\sigma^{2} - 1) \mathbf{I}\bigg)$, while the fifth column represents the generated samples obtained by denoising. The arrow at the bottom of Fig.~\ref{fig:sample} indicates the direction of time evolution in the stochastic differential equation, as per Eq.~\eqref{eq:rsde}. 

The order parameter or magnetization is defined as
\begin{equation}
    \left\langle M\right\rangle = \frac{1}{V}\left\langle \sum_{x}\phi(x)\right\rangle,
\end{equation}
where $V= N^d$ denotes the number of lattice sites. A comparison between HMC generated testing samples and DM generated samples is shown in Fig.~\ref{fig:brokenphase}, where we see consistency between the distributions containing 1024 samples each.

%%%%%%%%%%%%%%%%%%%%%%%%%%%%%%%%%%%%%%%%%%%%%%%%%%%%%%%%
%%%%%%%%%%%%%%%%%%%%%%%%%%%%%%%%%%%%%%%%%%%%%%%%%%%%%%%%

\subsection{Symmetric Phase}
\label{sec:sym_phase}
We now turn to the symmetric phase and take $\kappa=0.21$, below but near the critical value. In the thermodynamic limit, the phase transition can be characterized by a divergence in the connected two-point susceptibility. This susceptibility is a measure of how a system responds to small perturbations and is defined as
\begin{equation}
    \chi_2 = V\left( \langle M^2 \rangle - \langle M \rangle^2 \right).
\end{equation}
In addition, the Binder cumulant~\cite{Binder:1981zz} is a dimensionless quantity often used to identify phase transitions, defined as a ratio of moments of a distribution,
\begin{equation}
    U_L = 1- \frac{1}{3}\frac{\langle M^4 \rangle}{\langle M^2 \rangle^2}.
\end{equation}
It is sensitive to the shape of the distribution, i.e., the kurtosis of the order parameter, and is particularly useful for detecting and analyzing phase transitions in finite-size systems.

%%%%%%%%%%%%%%%%%%%%%%%%%%%%%%%%%%%%%%%%%%%%%%%%%%%%%%%%
\begin{figure}[!htbp]
\begin{center}
\includegraphics[width=0.6\textwidth]{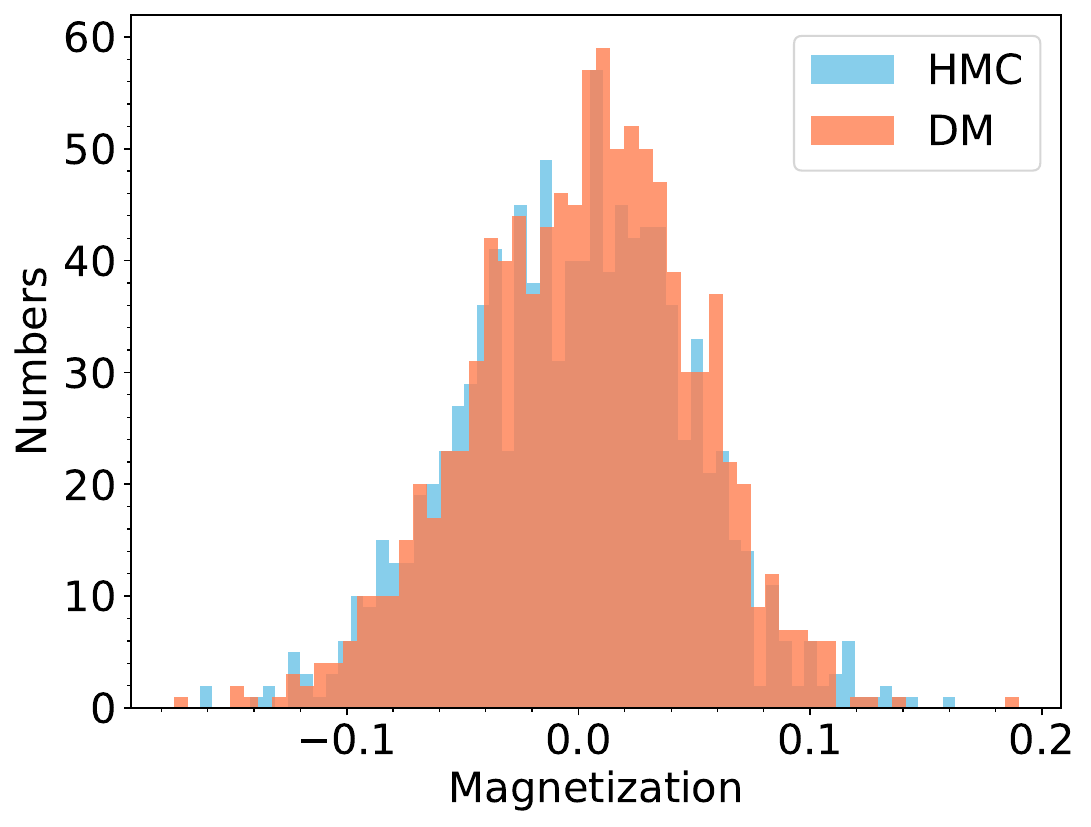}
\caption{Comparison of the distribution of the magnetization in the symmetric phase from the test data-set (HMC) and from the well-trained DM. The number of independent configurations is 1024 in both cases. 
}\label{fig:k021}
\end{center}
\end{figure}
%%%%%%%%%%%%%%%%%%%%%%%%%%%%%%%%%%%%%%%%%%%%%%%%%%%%%%%%%%%

%%%%%%%%%%%%%%%%%%%%%%%%%%%%%%%%%%%%%%%%%%%%%%%%%%%%%%%%%%%%%%%%%%%%%
\begin{table*}[!htbp]
\centering
\begin{tabular}{c|c|c|c}
\hline\hline
    data-set&$\langle M \rangle$ & $\chi_2$ & $U_L$ \\
\hline
    Training (HMC)& 0.0012$\pm$ 0.0007&  2.5160 $\pm$ 0.0457 & 0.1042 $\pm$ 0.0367\\
    Testing (HMC)& 0.0018 $\pm$ 0.0015& 2.4463 $\pm$ 0.1099 & -0.0198 $\pm$ 0.1035\\
    Generated (DM)& 0.0017$\pm$ 0.0015 &  2.4227 $\pm$ 0.1035 & 0.0484 $\pm$ 0.0959\\
\hline\hline
\end{tabular}
\caption{Comparison of observables calculated on different datasets with lower error bounds determined using the statistical jackknife method. The sample size is comparable for the bottom two rows and a factor of 5 larger for the training set (top row).
}
\label{tab:observables}
\end{table*}
%%%%%%%%%%%%%%%%%%%%%%%%%%%%%%%%%%%%%%%%%%%%%%%%%%%%%%%%%%%%%%%%%%%%%

In Fig.~\ref{fig:k021} we compare the distributions of the magnetization obtained independently from the trained DM and using HMC. Good consistency between the distributions is observed. To make this more quantitative we present in Table~\ref{tab:observables} the magnetization $\langle M\rangle$, two-point susceptibility $\chi_2$, and the Binder cumulant $U_L$ obtained using both the trained DM and with HMC. For the latter, expectation values are shown both for the training set and the testing set, with the former having a factor of five more configurations, which is reflected in the uncertainty. We conclude that the DM is capable of accurately capturing the statistical properties of the system, as reflected by these observables, and is in good agreement with HMC, a well-established method for sampling from the target distribution. {A study of the dependence on the size of the training set and the generated data set can be found in the Appendix~\ref{app:sta}.} We note that no accept-reject step has been applied in the DM.

\begin{figure}[!htbp]
\begin{center}
\includegraphics[width=0.6\textwidth]{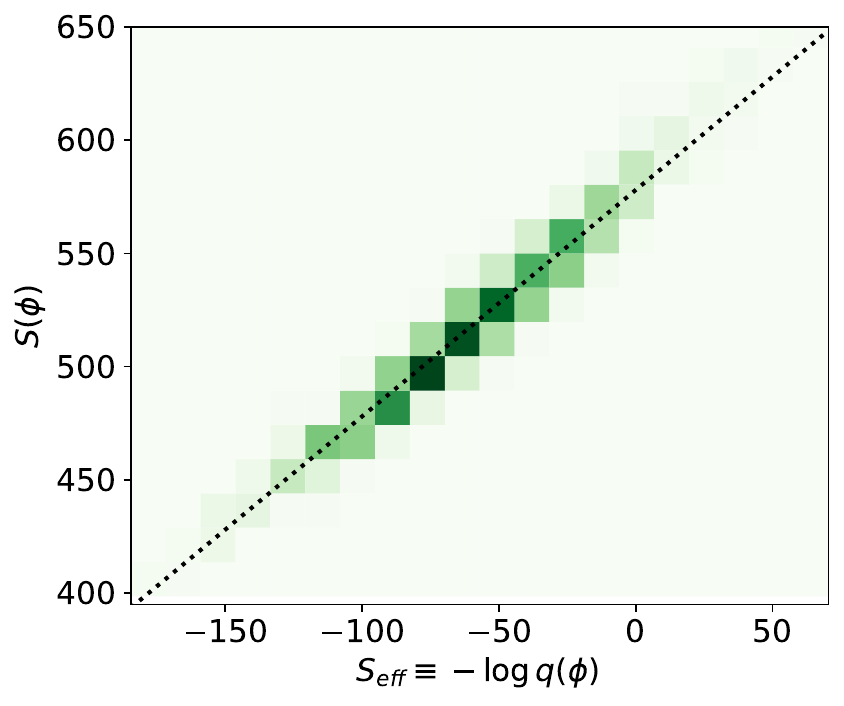}
\caption{Correlation between the effective action estimated from the trained DM using the probabilistic ODE and from the physical action on 1024 samples.}\label{fig:action}
\end{center}
\end{figure}

The likelihood computation method described in Section~\ref{subsec:continious} enables us to estimate the action for the field configurations generated by the trained DM. By comparing the action estimation from the trained DMs with the true action, see Eq.~(\ref{eq:phi4action}), we can confirm whether the DM has captured the underlying physics correctly. For this purpose, we first generate an ensemble of field configurations using the trained DM. We then compute the likelihoods, $q(\phi)$, for these configurations using the probability flow ODE approach. By applying the change-of-variable formula, we are able to compute the action values for these field configurations, as
\begin{equation}
    S_\text{eff} = -\log q(\phi),
\end{equation}
where the constant term $Z$ has been omitted. In Fig.~\ref{fig:action} we compare the evaluated effective action with the one obtained directly from the physical action, i.e., Eq.~\eqref{eq:phi4action}.  The black dashed line means the $x-$axis is proportional to the $y-$axis, and the green sites are action values estimated from the probabilistic ODE (on the $x-$axis) and calculated from the physical action (on the $y-$axis).

%%%%%%%%%%%%%%%%%%%%%%%%%%%%%%%%%%%%%%%%%%%%%%%%%%%%%%%%
%%%%%%%%%%%%%%%%%%%%%%%%%%%%%%%%%%%%%%%%%%%%%%%%%%%%%%%%%%%

The coefficient of determination, $R^2$, is a useful metric for quantitatively measuring how well the estimated action values, denoted as $S_{\rm eff}(\phi_i)$,  match the actual ones, denoted as $S(\phi_i)$. Here $i$ indicates the index in sampled configurations. We use
\begin{equation}
  R^2 = 1 - \frac{\sum_i\left[ S(\phi_i) - S_\text{eff}(\phi_i)\right]^2}{\sum_i \left[S(\phi_i) - \overline{S}(\phi)\right]},
\end{equation}
where $\overline{S}(\phi)$ is the average value for the actual action,
\begin{equation}
    \overline{S}(\phi) =\frac{1}{N}\sum_{i=1}^N S(\phi_i).
\end{equation}
It measures the proportion of variation in the dependent variable (in this case, the action) that is predictable from the independent variable (in this case, the DM-generated samples). $R^2$ values close to 1 indicate that the DM provides an accurate estimate of the action, suggesting that the trained DM has effectively learned the underlying physics of the scalar field theory correctly. 
In Fig.~\ref{fig:action}, with the default dataset size, i.e., 5120 samples for training, the $R^2$ value is 0.960 calculated on the 1024 generated configurations. 

{
\subsection{Autocorrelation Time}

Critical slowing down~\cite{Wolff:1989wq} emerges when simulations approach a critical point and the correlation length of the system diverges. This should be explored at $\kappa$ values close to the phase transition, which is located at $\kappa_c \simeq 0.27$ (for $L = 32, \lambda = 0.022$); see e.g.~Refs.~\cite{Pawlowski:2017rhn,Pawlowski:2018qxs,Bachtis:2020ajb}.

Critical slowing down can be studied using the autocorrelation function of an observable $O$, defined as
\begin{equation}
    C_{O}(t) = \left\langle (O_{t_0} - \left\langle O_{t_0} \right\rangle)(O_{t_0+t} - \left\langle O_{t_0+t} \right\rangle) \right\rangle 
            = \left\langle O_{t_0}O_{t_0+t} \right\rangle - \left\langle O_{t_0} \right\rangle\left\langle O_{t_0+t} \right\rangle,
\end{equation}
where $t_0$ is the starting time and $t$ denotes the time along the Markov chain. In an exponential fit, $C_O(t)\sim \exp\{ -t/t_c\}$, the characteristic time $t_c$ should scale as a power of the correlation length $\xi$ around the critical point, i.e., $t_c\sim \xi^z$ (the correlation length $\xi$ should not be confused with the time variable in the forward SDE). The normalized autocorrelation functions, $\bar{C}_O(t) \equiv C_O(t)/C_O(0)$ can be used for a clearer comparison.

%%%%%%%%%%%%%%%%%%%%%%%%%%%%%%%%%%%%%%%%%%%%%%%%%%%%%%%%
\begin{figure}[!htbp]
\begin{center}
\includegraphics[width=0.7\textwidth]{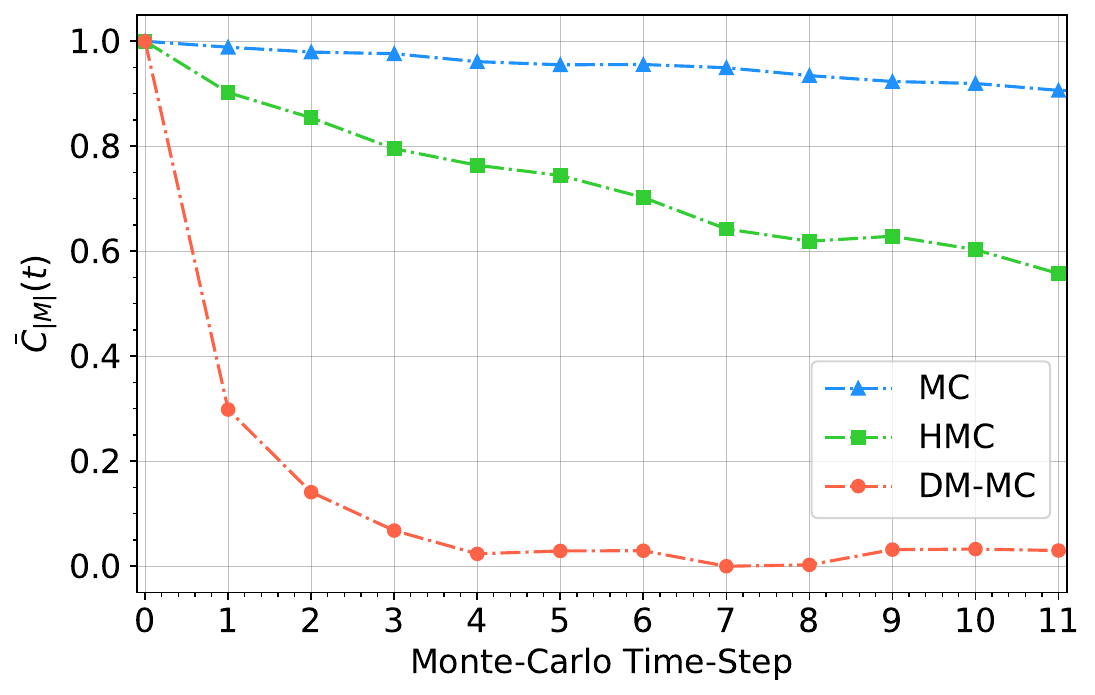}
\caption{Comparison of the normalized autocorrelation function of $|M|$ for the Metropolis-Hastings MC algorithm, HMC, and DM-based MC, using 1024 samples at $\kappa = 0.27, \lambda  = 0.22$ and $L = 32$}.
\label{fig:autocor}
\end{center}
\end{figure}
%%%%%%%%%%%%%%%%%%%%%%%%%%%%%%%%%%%%%%%%%%%%%%%%%%%%%%%%%%%

We investigate this for the ``absolute value'' of the magnetization, defined as 
\begin{equation}
|M| = \frac{1}{V} \left\langle  \sum_x \left|\phi(x) \right|\right\rangle,
\end{equation}
and compare various updates. Because the likelihood of samples can be computed, as shown in the previous section, we can use the Metropolis-Hastings (MH) algorithm to construct the asymptotically exact Markov chain sampler with the well-trained DM. We refer to this as the DM-based Monte-Carlo method, following the convention of flow-based models. The acceptance criterion is corrected as 
\begin{equation}
p_\text{accept}(\phi_\text{proposal}\mid \phi^{i-1}) = \text{min} \left ( 1, \frac{q(\phi^{i-1})}{p(\phi^{i-1})}\frac{p(\phi_\text{proposal})}{q(\phi_\text{proposal})} \right ),
\end{equation}
where $\phi^{i-1}$ is the previous configuration and $\phi_\text{proposal}$ is sampled from the well-trained DM. Further, we can estimate the efficiency gain of the DM by comparing the behavior of the autocorrelation function $C_{|M|}(t)$ with the one obtained using a Metropolis-Hastings Monte-Carlo (MC) and the HMC algorithm on 1024 configurations, as shown in Fig.~\ref{fig:autocor}. {A more comprehensive discussion on the dependence of the acceptance rate on the parameters and the number of training epochs can be found in Appendix~\ref{app:acc}}.

Beside, one can compute the integrated autocorrelation time~\cite{Schaefer:2010hu,Pawlowski:2018qxs,DelDebbio:2021qwf} for a more quantitative comparision. It is defined as
\begin{equation}
    \tau_{O,{\mathrm{int}}}={\frac{1}{2}}+{\frac{1}{C_{O}(0)}}\sum_{t=1}^{t_\text{max}}C_{O}(t).
\end{equation}
For the algorithms discussed, integrated autocorrelation times are $ \tau_{|M|,{\mathrm{int}}}\text{(MC)} = 79.984$, $\tau_{|M|,{\mathrm{int}}}\text{(HMC)} = 41.354$, and $\tau_{|M|,{\mathrm{int}}}\text{(DM-MC)} = 2.360$, where we calculated the autocorrelation time with $t_\text{max} = 100$ MC trajectories on 1024 configurations. To get a more precise estimation with uncertainties, one can also choose the automatic windowing procedure in the computation~\cite{Wolff:2003sm,Joswig:2022qfe}. These comparisons demonstrate that our proposed method has the potential to significantly suppress autocorrelations along a Markov chain.}

%%%%%%% SECTION %%%%%%%%%%%%%%%%%

\section{Conclusion and Outlook}
\label{sec:sum}

\begin{figure}[hbtp!]
    \centering
    \includegraphics[width = 0.7 \textwidth]{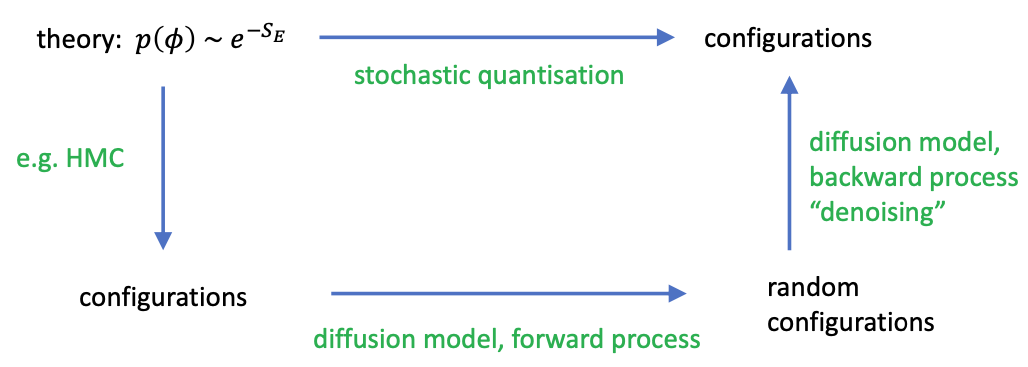}
    \caption{Schematic overview of the relation between stochastic quantisation and diffusion models in the case of lattice field theory. }
    \label{fig:flow-chart}
\end{figure}

In this work, we proposed a new method to generate quantum field configurations using diffusion models, popular in the ML community. We highlighted the connection with stochastic quantization (SQ), an approach to quantize field theories based on a stochastic process in a fictitious time direction. In SQ the drift term in the stochastic process is known, as it is derived from the probability distribution one wants to sample from, which is determined by the quantum theory under consideration. In DMs the drift term is not known but is learned from data in a forward stochastic process. After estimating the drift term, the trained DM is run in the opposite direction and independent configurations are created from noise, the ``denoising'' process. This latter part is akin to the dynamics in SQ, albeit with a learned and time-dependent drift term. {In Figure~\ref{fig:flow-chart}, we summarize the relation between stochastic quantisation and diffusion models schematically.}

%%%%%%%%%%%%%%%%%%%%%%%%%%%%%%%%%%%%%%%%%%%%%%%%%%%%%%%%%%%%%%%%%%%%%%%%%%%%%%%%%%%%%%%%%%%%%%%%%%%%%%%%%%%%%%%%%%
%%%%%%%%%%%%%%%%%%%%%%%%%%%%%%%%%%%%%%%%%%%%%%%%%%%%%%%%%%%%%%%%%%%%%%%%%%%%%%%%%%%%%%%%%%%%%%%%%%%%%%%%%%%%%%%%%%

We have demonstrated the approach in a simple toy model and in a two-dimensional scalar field theory with a $\phi^4$ interaction, both in the symmetric and in the broken phase. We have shown that the DM can serve as a global sampler to assist methods based on the Monte Carlo principle. We note that one can include an accept-reject step at the end of each trajectory. We have found in our numerical results that the well-trained DM can successfully reproduce configurations in both phases and that autocorrelation times are reduced along a Markov chain. This improvement in sampling efficiency can be beneficial for studying lattice field theories and should be analyzed further. 

There are various directions to explore in the future. The forward (``smoothening") and backward (``denoising") processes are reminiscent of (inverse) renormalization group transformations and it would be interesting to make this precise.   
Because the action can be estimated relatively accurately, it is feasible to train a DM without preparing a training data-set, which needs to be investigated.
The connection between DMs and SQ offers new perspectives. It is well known how to implement SQ for non-abelian gauge theories; hence it should be possible to formulate DMs for the non-abelian case. An intriguing direction is the following: imagine one has generated a set of configurations for a quantum field theory with fermions, such as QCD or the Schwinger model. Since the fermions are ``integrated out'', the configurations are given in terms of the gauge fields only. By using these configurations as the starting point of a DM, one can learn an effective stochastic process which incorporates the effect of the fermions implicitly. This may be used to create additional configurations in a numerically ``cheap" manner. An accept-reject step based on the underlying theory can be included, which would require the evaluation of the fermion determinant. Finally SQ can also be used for theories with a sign or complex action problem, and hence a non-positive definite distribution. This is commonly known as complex Langevin (CL) dynamics. It is known that CL may become less reliable along long trajectories in the stochastic process. Combining CL with DMs, one may consider fairly short trajectories, but subsequently use the resulting configurations to train the DM to generate additional configurations to increase statistics.

\acknowledgments

We thank Profs.\ Tetsuo Hatsuda, Shuzhe Shi and Xu-Guang Huang for helpful discussions. 
We thank ECT* and the ExtreMe Matter Institute EMMI at GSI, Darmstadt, for support during the ECT*/EMMI workshop {\em Machine learning for lattice field theory and beyond} in June 2023 during the preparation of this paper.
The work is supported by (i) the CUHK-Shenzhen university development fund under grant No. UDF01003041 and the BMBF funded KISS consortium (05D23RI1) in the ErUM-Data action plan (K.Z.), (ii) the AI grant of SAMSON AG, Frankfurt (K.Z.\ and L.W.), (iii) Xidian-FIAS International Joint Research Center (L.W), (iv) STFC Consolidated Grant ST/T000813/1 (G.A.). K.Z.\ also thanks the donation of NVIDIA GPUs from NVIDIA Corporation. L.W. \ also thanks the Natural Science Foundation of China  (Grant No.12147101) for supporting the computation resources when visiting in Fudan University.

\paragraph{Note added.} Recent related work introduces the stochastic process into the hybrid Monte-Carlo algorithm~\cite{Robnik:2023pgt} and explores the correspondence between the exact renormalizing group (ERG) and DMs based upon the heat equation~\cite{Cotler:2023lem}.

\appendix

\section{U-Net Architecture}\label{app:unet}

\begin{figure}[!htbp]
\begin{center}
\includegraphics[width=0.9\textwidth]{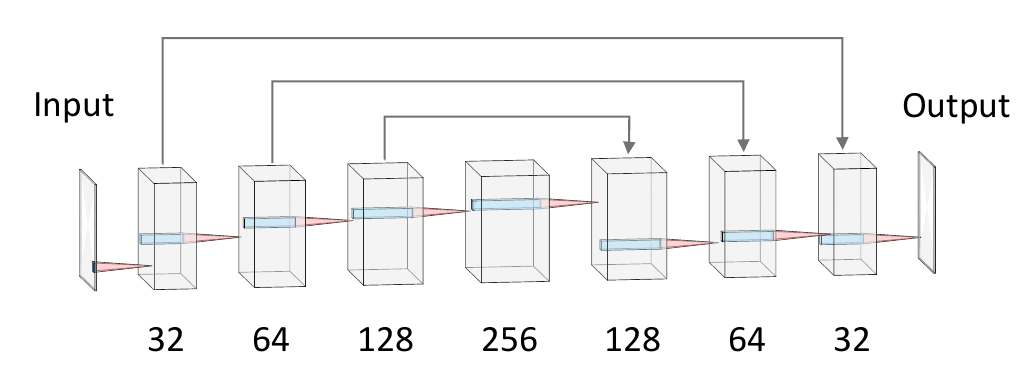}
\caption{Sketch of the U-Net Architecture, where the gray lines with arrows indicate skip connections between encoding paths and decoding paths. In this study, the default shape of the input is set as $(\cdot,1,32,32)$, where the first dimension is the batch size, the second indicates the channel, and the final two represent dimensionalities of the field configuration.}\label{fig:unet}
\end{center}
\end{figure}

U-Net architectures are utilized to present $s_\mathbf{\theta}(\phi,t)$ with a time-embedding input. This architecture is widely used in tasks that require semantic segmentation due to its ability to capture both local and global information in an image. The model is initialized with a function that calculates the standard deviation of the perturbation kernel, a list of channels for feature maps at each resolution, and an embedding dimension for Gaussian random feature embeddings. The model consists of an \textbf{encoding path} (contracting path) and a \textbf{decoding path} (expansive path), which are characteristic of the U-Net architecture, as Fig.~\ref{fig:unet} shows. 

%%%%%%%%%%%%%%%%%%%%%%%%%%%%%%%%%%%%%%%%%%%%%%%%%%%%%%%%
%%%%%%%%%%%%%%%%%%%%%%%%%%%%%%%%%%%%%%%%%%%%%%%%%%%%%%%%%%%

The \textbf{encoding path} consists of four convolutional layers (conv1 to conv4), each followed by a dense layer (dense1 to dense4), and a group normalization layer (gnorm1 to gnorm4). The convolutional layers progressively increase the number of channels from 1 to 32, 64, 128, 256. The dense layers incorporate information from the time embedding into the feature maps, and the group normalization layers normalize the feature maps across the channels.

The \textbf{decoding path} consists of four transposed convolutional layers (tconv4 to tconv1), each followed by a dense layer (dense5 to dense7), and a group normalization layer (tgnorm4 to tgnorm2). The transposed convolutional layers progressively decrease the number of channels back to 1. The dense layers incorporate information from the time embedding into the feature maps, and the group normalization layers normalize the feature maps across the channels. The decoding path also includes skip connections from the encoding path, which help the model to better localize and learn representations with less loss of spatial information.

The model uses the Swish activation function, which is a smooth, non-monotonic function that has been found to work better than ReLU in deeper models. In the forward pass, the model first obtains the Gaussian random feature embedding for the input time $t$. It then passes the input $\phi$ through the encoding path, where it applies the convolutional, dense, group normalization, and activation layers sequentially. It then passes the output of the encoding path through the decoding path, where it applies the transposed convolutional, dense, group normalization, and activation layers sequentially, while also adding the skip connections from the encoding path. Finally, it normalizes the output by the standard deviation of the perturbation kernel at time $t$ and returns it.

\section{Skilling-Hutchinson Estimator}\label{app:she}

The divergence function in Eq.~\eqref{eq:jacobian} can be hard to evaluate in practice for general vector-valued functions $f$, but we can use an unbiased estimator, named Skilling-Hutchinson estimator~\cite{grathwohl2018scalable}, to approximate the trace. Let $\boldsymbol \epsilon \sim \mathcal{N}(\mathbf{0}, \mathbf{I})$. The Skilling-Hutchinson estimator is based on the fact that
\begin{equation}
\mathbf{\nabla}\cdot f(\phi) = \mathbb{E}_{\boldsymbol\epsilon \sim \mathcal{N}(\mathbf{0}, \mathbf{I})}\left[\boldsymbol\epsilon^\intercal  J_f(\phi) \boldsymbol\epsilon\right].
\end{equation}
Therefore, we can simply sample a random vector $\boldsymbol \epsilon \sim \mathcal{N}(\mathbf{0}, \mathbf{I})$, and then use $\boldsymbol \epsilon^\intercal J_f(\phi) \boldsymbol \epsilon$ to estimate the divergence of $f(\phi)$. This estimator only requires computing the Jacobian-vector product $J_f(\phi)\boldsymbol \epsilon$, which is typically efficient.

As a result, for our probability flow ODE, we can compute the (log) data likelihood with the following,
\begin{equation}
\log p_0(\phi(0)) = \log p_T(\phi(T)) -\int_0^T \frac{d[\sigma^2(t)]}{dt} \mathbf{\nabla}\cdot s_\theta(\phi(t), t) dt.
\end{equation}
With the Skilling-Hutchinson estimator, we can compute the divergence via
\begin{equation}
\mathbf{\nabla}\cdot s_\theta(\phi(t), t) = \mathbb{E}_{\boldsymbol\epsilon \sim \mathcal{N}(\mathbf{0}, \mathbf{I})}\left[\boldsymbol\epsilon^\intercal  J_{s_\theta}(\phi(t), t) \boldsymbol\epsilon\right].
\end{equation}
Subsequently, we can compute the integral using numerical integrators. This provides us with an unbiased estimate of the true likelihood of the data, and we can make this estimate increasingly accurate by running the calculation multiple times and taking the average. The numerical integrator requires $\phi(t)$ as a function of $t$, which can be obtained by the probability flow ODE sampler. In our calculations, we tested different sampling sizes ranging from 1 to 100 and found that it converges when the size is larger than 10. Therefore, we chose the sample size $\boldsymbol\epsilon$ as 10 to ensure the accuracy of the estimation.

%%%%%%%%%%%%%%%%%%%%%%%%%%%%%%%%%%%%%%%%%%%%%%%%%%%%%%%%%%%%%
\begin{figure}[t]
    \centering
    \includegraphics[width = 0.8 \textwidth]{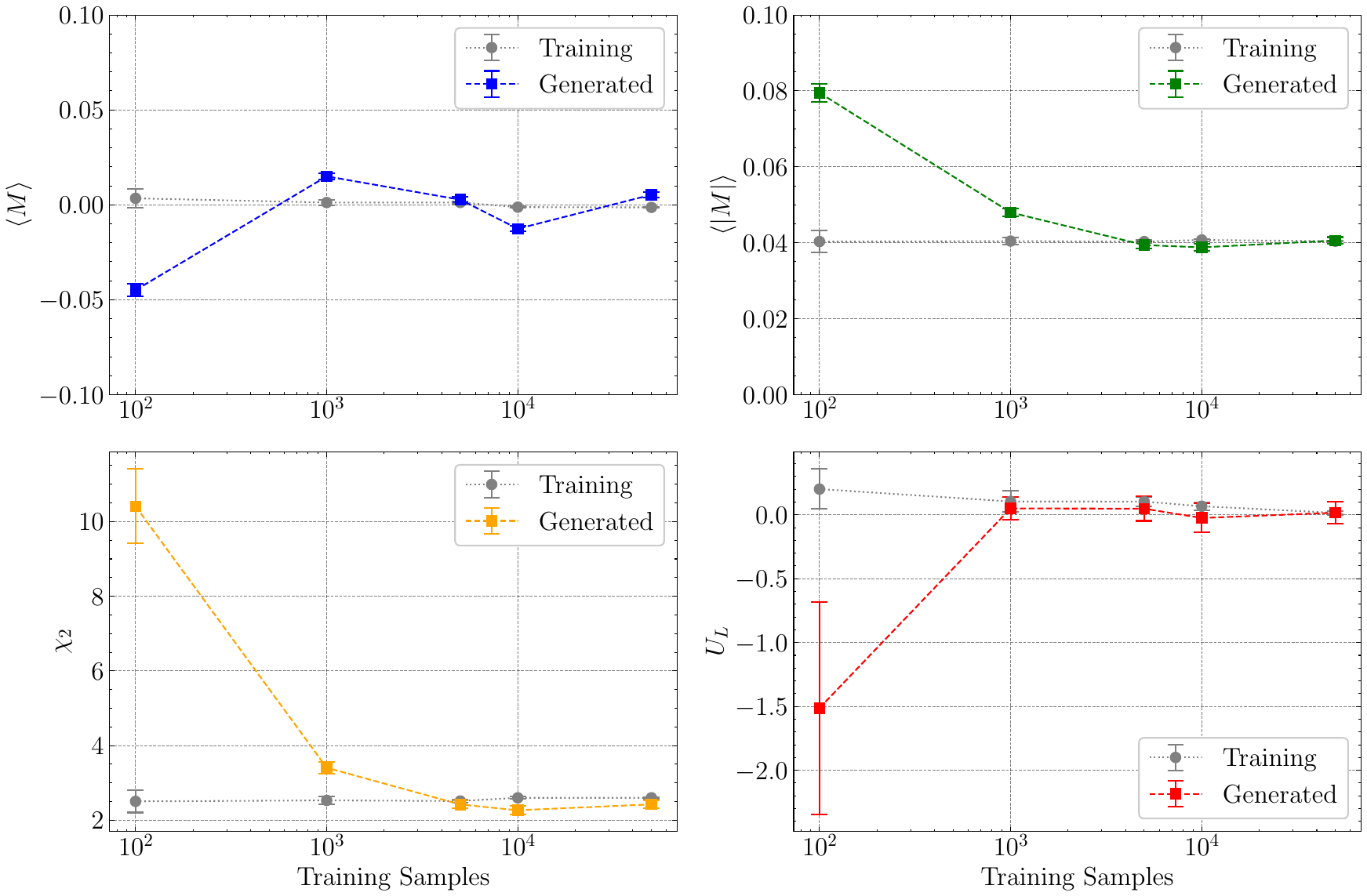}
    \caption{Dependence of four observables and their uncertainties on the size of the training dataset within the symmetric phase ($\kappa = 0.21, \lambda=0.022$), using a fixed number of 1024 generated configurations.}
    \label{fig:training}
\end{figure}
%%%%%%%%%%%%%%%%%%%%%%%%%%%%%%%%%%%%%%%%%%%%%%%%%%%%%%%%%%%%%
%%%%%%%%%%%%%%%%%%%%%%%%%%%%%%%%%%%%%%%%%%%%%%%%%%%%%%%%%%%%%
\begin{figure}[t]
    \centering
    \includegraphics[width = 0.8 \textwidth]{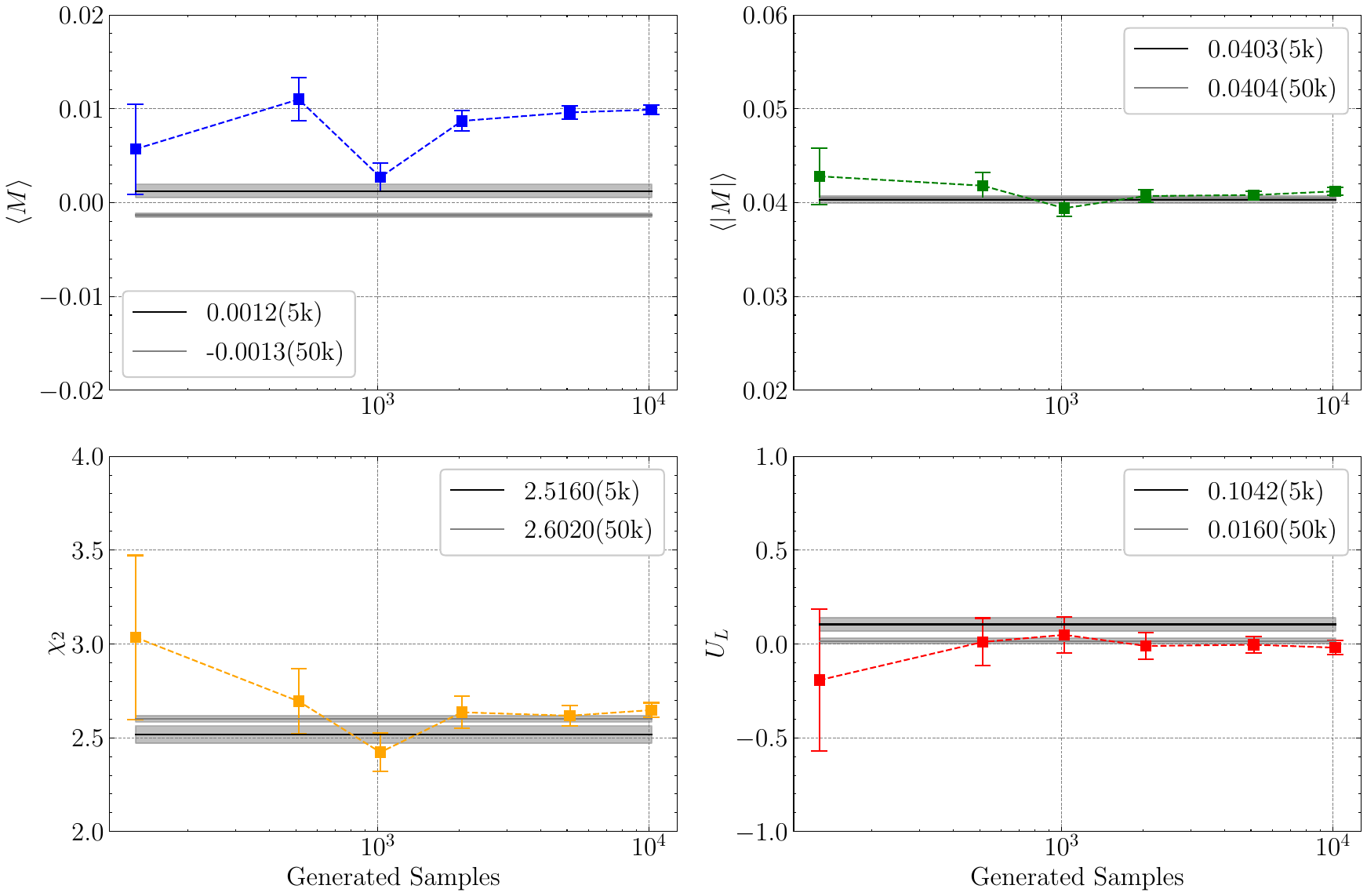}
    \caption{Dependence of observables on the number of generated samples obtained from a DM trained on a dataset of 5120 configurations within the symmetric phase ($\kappa = 0.21, \lambda=0.022$). The DM results are compared with predictions from the training set and with those from a larger unused dataset with 51200 configurations. }
    \label{fig:generated}
\end{figure}
%%%%%%%%%%%%%%%%%%%%%%%%%%%%%%%%%%%%%%%%%%%%%%%%%%%%%%%%%%%%%
{

\section{Statistics Examination}\label{app:sta}

In Figure~\ref{fig:training} (with the same physical parameter setup as in Sec.~\ref{sec:sym_phase}), we demonstrate the dependence of four key physical observables including their uncertainties on the size of the training datasets, while the size of the generated dataset remains constant (1024 configurations). The illustration uses gray markers to denote values derived from various training datasets, while colored markers represent values obtained from the generated datasets. The analysis reveals a trend towards convergence in the estimations of observables as the size of the training dataset increases, even with a relatively small generated dataset size. 

In Figure~\ref{fig:generated}, we present the dependence of the observables on the number of generated samples obtained from a DM trained on a dataset of 5120 configurations. The observables' values calculated on the training dataset are traced with black lines, contrasting with gray lines that represent calculations over a larger, unused dataset (51200 configurations). The results show that observables from the generated datasets tend to align with those from the larger dataset as their size increases. 

Although there is a noticeable deviation in the $\langle M \rangle$ estimations towards the training dataset values, it can be conceivably improved by introducing the $\mathrm{Z}_2$ symmetry into the training data-sets explicitly, e.g., data augmentation with flipping all configurations.

\section{Acceptance Rate}\label{app:acc}

We investigate the dependence of the acceptance rate on the value of the hopping parameter, the system size and the number of training epochs. 

%%%%%%%%%%%%%%%%%%%%%%%%%%%%%%%%%%%%%%%%%%%%%%%%%%%%%%%%%%%%%
\begin{figure}[!htbp]
    \centering
    \includegraphics[width = 0.5 \textwidth]{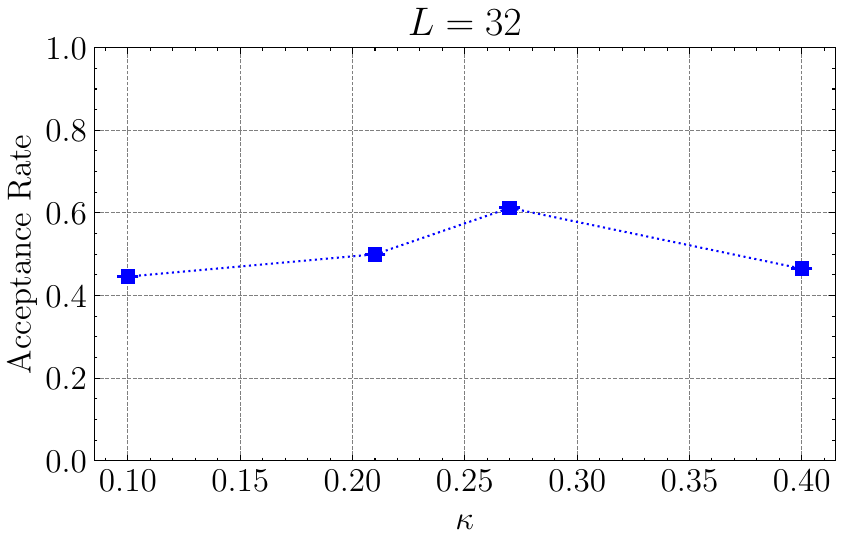}
    \caption{Dependence of the acceptance rate on the hopping parameter $\kappa$ on a $32\times 32$ lattice. 
    }
    \label{app_fig:acc_k}
\end{figure}
%%%%%%%%%%%%%%%%%%%%%%%%%%%%%%%%%%%%%%%%%%%%%%%%%%%%%%%%%%%%%

We have calculated the dependence of the acceptance rate on the hopping parameter, specifically at \(\kappa = 0.1, 0.21, 0.27, 0.4\), using an identical setup as in the main part of the paper, i.e.\ with 250 training epochs, \(T=100\) MC time-steps and 1024 configurations. In Figure~\ref{app_fig:acc_k}, we observe that the acceptance rate lies between 0.4 and 0.6. This observation indicates that the efficacy of the DM-MC method remains robust against the variations in interaction couplings, demonstrating its utility as an efficient sampling technique across a broad range of parameter values.

%%%%%%%%%%%%%%%%%%%%%%%%%%%%%%%%%%%%%%%%%%%%%%%%%%%%%%%%%%%%%
\begin{figure}[!htbp]
    \centering
    \includegraphics[width = 0.5 \textwidth]{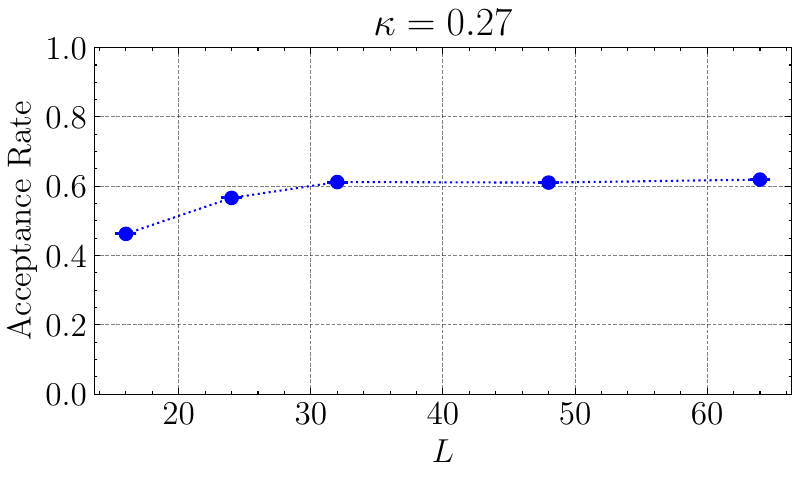}
    \caption{Dependence of the acceptance rate on the linear lattice size at a fixed hopping parameter \(\kappa = 0.27\).}
    \label{app_fig:acc_L}
\end{figure}
%%%%%%%%%%%%%%%%%%%%%%%%%%%%%%%%%%%%%%%%%%%%%%%%%%%%%%%%%%%%%

Next we vary the linear size of the lattice, \(L = 16, 24, 32, 48, 64\), at fixed  \(\kappa = 0.27\), using an otherwise identical setup as above. In Figure~\ref{app_fig:acc_k}, we observe that the acceptance rate converges to around 0.6 for larger sizes, suggesting that the efficacy of the DM-MC will not be affected by the system size.

%%%%%%%%%%%%%%%%%%%%%%%%%%%%%%%%%%%%%%%%%%%%%%%%%%%%%%%%%%%%%
\begin{figure}[hbtp!]
    \centering
    \includegraphics[width = 0.5 \textwidth]{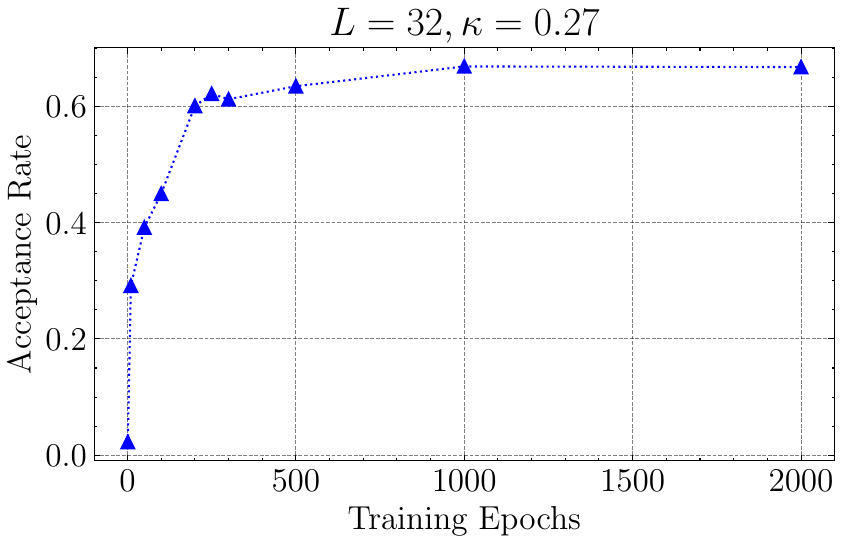}
    \caption{Dependence of the acceptance rate on the number of epochs used for training, at $\kappa = 0.27, L = 32, \lambda = 0.022$.}
    \label{app_fig:acc_epochs}
\end{figure}
%%%%%%%%%%%%%%%%%%%%%%%%%%%%%%%%%%%%%%%%%%%%%%%%%%%%%%%%%%%%%

The dependence of the acceptance rate of DM-MC on the number of training epochs is shown in Figure~\ref{app_fig:acc_epochs} for parameter values closer to the critical point. It shows that the acceptance rate increases rapidly within the first 200 epochs and converges slowly over the remaining 1800 epochs. Although our choice of 250 epochs used in the main part of the paper yields an acceptance rate slightly below the asymptotic value, it is indeed an acceptable balance between training cost and performance. 

Concerning the dependence of the training cost on the various parameters, we have found that 
\begin{itemize}
\item for different $\kappa$ values on a $32\times 32$ lattice the training cost is almost constant (at around 6.54 seconds per epoch on our infrastructure);
\item increasing the volume results in a longer training time; for linear size \(L =\) 16, 24, 32, 48, 64, the training cost is \(5.80, 6.19, 6.54, 6.97, 8.63\) seconds per epoch on our infrastructure. A rough fit then yields a training time of $L^{0.43} + 2.31$ seconds per epoch.
\end{itemize}
}

\bibliographystyle{JHEP}
\bibliography{dm4lqft_v2}

\providecommand{\href}[2]{#2}\begingroup\raggedright\begin{thebibliography}{10}

\bibitem{Knechtli:2017sna}
F.~Knechtli, M.~G\"unther and M.~Peardon, \emph{{Lattice Quantum
  Chromodynamics: Practical Essentials}}, SpringerBriefs in Physics, Springer
  (2017),
  \href{https://doi.org/10.1007/978-94-024-0999-4}{10.1007/978-94-024-0999-4}.

\bibitem{Wolff:1989wq}
U.~Wolff, \emph{{CRITICAL SLOWING DOWN}},
  \href{https://doi.org/10.1016/0920-5632(90)90224-I}{\emph{Nucl. Phys. B Proc.
  Suppl.} {\bfseries 17} (1990) 93}.

\bibitem{tomczak:2022deep}
J.M.~Tomczak, \emph{Deep Generative Modeling}, {Springer International
  Publishing}, {Cham} (2022),
  \href{https://doi.org/10.1007/978-3-030-93158-2}{10.1007/978-3-030-93158-2}.

\bibitem{Boyda:2022nmh}
D.~Boyda et~al., \emph{{Applications of Machine Learning to Lattice Quantum
  Field Theory}},  in \emph{{Snowmass 2021}}, 2, 2022
  [\href{https://arxiv.org/abs/2202.05838}{{\ttfamily 2202.05838}}].

\bibitem{Zhou:2018ill}
K.~Zhou, G.~Endr{\H o}di, L.-G.~Pang and H.~St{\"o}cker, \emph{Regressive and
  generative neural networks for scalar field theory},
  \href{https://doi.org/10.1103/PhysRevD.100.011501}{\emph{Phys. Rev. D}
  {\bfseries 100} (2019) 011501}.

\bibitem{Pawlowski:2018qxs}
J.M.~Pawlowski and J.M.~Urban, \emph{{Reducing Autocorrelation Times in Lattice
  Simulations with Generative Adversarial Networks}},
  \href{https://doi.org/10.1088/2632-2153/abae73}{\emph{Mach. Learn. Sci.
  Tech.} {\bfseries 1} (2020) 045011}
  [\href{https://arxiv.org/abs/1811.03533}{{\ttfamily 1811.03533}}].

\bibitem{Albergo:2019eim}
M.S.~Albergo, G.~Kanwar and P.E.~Shanahan, \emph{Flow-based generative models
  for {{Markov}} chain {{Monte Carlo}} in lattice field theory},
  \href{https://doi.org/10.1103/PhysRevD.100.034515}{\emph{Phys. Rev. D}
  {\bfseries 100} (2019) 034515}
  [\href{https://arxiv.org/abs/1904.12072}{{\ttfamily 1904.12072}}].

\bibitem{Kanwar:2020xzo}
G.~Kanwar, M.S.~Albergo, D.~Boyda, K.~Cranmer, D.C.~Hackett, S.~Racani{\`e}re
  et~al., \emph{Equivariant {{Flow-Based Sampling}} for {{Lattice Gauge
  Theory}}}, \href{https://doi.org/10.1103/PhysRevLett.125.121601}{\emph{Phys.
  Rev. Lett.} {\bfseries 125} (2020) 121601}
  [\href{https://arxiv.org/abs/2003.06413}{{\ttfamily 2003.06413}}].

\bibitem{Nicoli:2020njz}
K.A.~Nicoli, C.J.~Anders, L.~Funcke, T.~Hartung, K.~Jansen, P.~Kessel et~al.,
  \emph{Estimation of {{Thermodynamic Observables}} in {{Lattice Field
  Theories}} with {{Deep Generative Models}}},
  \href{https://doi.org/10.1103/PhysRevLett.126.032001}{\emph{Phys. Rev. Lett.}
  {\bfseries 126} (2021) 032001}
  [\href{https://arxiv.org/abs/2007.07115}{{\ttfamily 2007.07115}}].

\bibitem{Albergo:2021bna}
M.S.~Albergo, G.~Kanwar, S.~Racani{\`e}re, D.J.~Rezende, J.M.~Urban, D.~Boyda
  et~al., \emph{Flow-based sampling for fermionic lattice field theories},
  \href{https://doi.org/10.1103/PhysRevD.104.114507}{\emph{Phys. Rev. D}
  {\bfseries 104} (2021) 114507}
  [\href{https://arxiv.org/abs/2106.05934}{{\ttfamily 2106.05934}}].

\bibitem{Albergo:2021vyo}
M.S.~Albergo, D.~Boyda, D.C.~Hackett, G.~Kanwar, K.~Cranmer, S.~Racani{\`e}re
  et~al., \emph{Introduction to {{Normalizing Flows}} for {{Lattice Field
  Theory}}}, {\emph{arXiv:2101.08176 [hep-lat]} (2021) }
  [\href{https://arxiv.org/abs/2101.08176}{{\ttfamily 2101.08176}}].

\bibitem{DelDebbio:2021qwf}
L.~Del~Debbio, J.~Marsh~Rossney and M.~Wilson, \emph{Efficient modeling of
  trivializing maps for lattice $\phi^4$ theory using normalizing flows: {{A}}
  first look at scalability},
  \href{https://doi.org/10.1103/PhysRevD.104.094507}{\emph{Phys. Rev. D}
  {\bfseries 104} (2021) 094507}
  [\href{https://arxiv.org/abs/2105.12481}{{\ttfamily 2105.12481}}].

\bibitem{Hackett:2021idh}
D.C.~Hackett, C.-C.~Hsieh, M.S.~Albergo, D.~Boyda, J.-W.~Chen, K.-F.~Chen
  et~al., \emph{{Flow-based sampling for multimodal distributions in lattice
  field theory}},  \href{https://arxiv.org/abs/2107.00734}{{\ttfamily
  2107.00734}}.

\bibitem{Albergo:2022qfi}
M.S.~Albergo, D.~Boyda, K.~Cranmer, D.C.~Hackett, G.~Kanwar, S.~Racani{\`e}re
  et~al., \emph{Flow-based sampling in the lattice {{Schwinger}} model at
  criticality}, \href{https://doi.org/10.1103/PhysRevD.106.014514}{\emph{Phys.
  Rev. D} {\bfseries 106} (2022) 014514}
  [\href{https://arxiv.org/abs/2202.11712}{{\ttfamily 2202.11712}}].

\bibitem{Bacchio:2022vje}
S.~Bacchio, P.~Kessel, S.~Schaefer and L.~Vaitl, \emph{Learning {{Trivializing
  Gradient Flows}} for {{Lattice Gauge Theories}}},
  \href{https://doi.org/10.1103/PhysRevD.107.L051504}{\emph{Phys. Rev. D}
  {\bfseries 107} (2023) L051504}
  [\href{https://arxiv.org/abs/2212.08469}{{\ttfamily 2212.08469}}].

\bibitem{Caselle:2022acb}
M.~Caselle, E.~Cellini, A.~Nada and M.~Panero, \emph{Stochastic normalizing
  flows as non-equilibrium transformations},
  \href{https://doi.org/10.1007/JHEP07(2022)015}{\emph{JHEP} {\bfseries 07}
  (2022) 015} [\href{https://arxiv.org/abs/2201.08862}{{\ttfamily
  2201.08862}}].

\bibitem{Chen:2022ytr}
S.~Chen, O.~Savchuk, S.~Zheng, B.~Chen, H.~Stoecker, L.~Wang et~al.,
  \emph{{Fourier-flow model generating Feynman paths}},
  \href{https://doi.org/10.1103/PhysRevD.107.056001}{\emph{Phys. Rev. D}
  {\bfseries 107} (2023) 056001}
  [\href{https://arxiv.org/abs/2211.03470}{{\ttfamily 2211.03470}}].

\bibitem{Gerdes:2022eve}
M.~Gerdes, P.~de~Haan, C.~Rainone, R.~Bondesan and M.C.N.~Cheng,
  \emph{{Learning Lattice Quantum Field Theories with Equivariant Continuous
  Flows}},  \href{https://arxiv.org/abs/2207.00283}{{\ttfamily 2207.00283}}.

\bibitem{Albandea:2023wgd}
D.~Albandea, L.~Del~Debbio, P.~Hern\'andez, R.~Kenway, J.~Marsh~Rossney and
  A.~Ramos, \emph{{Learning trivializing flows}},
  \href{https://doi.org/10.1140/epjc/s10052-023-11838-8}{\emph{Eur. Phys. J. C}
  {\bfseries 83} (2023) 676}
  [\href{https://arxiv.org/abs/2302.08408}{{\ttfamily 2302.08408}}].

\bibitem{Singha:2023cql}
A.~Singha, D.~Chakrabarti and V.~Arora, \emph{Conditional normalizing flow for
  {{Markov}} chain {{Monte Carlo}} sampling in the critical region of lattice
  field theory}, \href{https://doi.org/10.1103/PhysRevD.107.014512}{\emph{Phys.
  Rev. D} {\bfseries 107} (2023) 014512}.

\bibitem{Abbott:2022zsh}
R.~Abbott et~al., \emph{{Aspects of scaling and scalability for flow-based
  sampling of lattice QCD}},
  \href{https://arxiv.org/abs/2211.07541}{{\ttfamily 2211.07541}}.

\bibitem{Nicoli:2023qsl}
K.A.~Nicoli, C.J.~Anders, T.~Hartung, K.~Jansen, P.~Kessel and S.~Nakajima,
  \emph{{Detecting and Mitigating Mode-Collapse for Flow-based Sampling of
  Lattice Field Theories}},  \href{https://arxiv.org/abs/2302.14082}{{\ttfamily
  2302.14082}}.

\bibitem{Wang:2020hji}
L.~Wang, Y.~Jiang, L.~He and K.~Zhou, \emph{{Continuous-Mixture Autoregressive
  Networks Learning the Kosterlitz-Thouless Transition}},
  \href{https://doi.org/10.1088/0256-307X/39/12/120502}{\emph{Chin. Phys.
  Lett.} {\bfseries 39} (2022) 120502}
  [\href{https://arxiv.org/abs/2005.04857}{{\ttfamily 2005.04857}}].

\bibitem{Luo:2023opo}
D.~Luo, Z.~Chen, K.~Hu, Z.~Zhao, V.M.~Hur and B.K.~Clark,
  \emph{{Gauge-invariant and anyonic-symmetric autoregressive neural network
  for quantum lattice models}},
  \href{https://doi.org/10.1103/PhysRevResearch.5.013216}{\emph{Phys. Rev.
  Res.} {\bfseries 5} (2023) 013216}.

\bibitem{Favoni:2020reg}
M.~Favoni, A.~Ipp, D.I.~M{\"u}ller and D.~Schuh, \emph{Lattice {{Gauge
  Equivariant Convolutional Neural Networks}}},
  \href{https://doi.org/10.1103/PhysRevLett.128.032003}{\emph{Phys. Rev. Lett.}
  {\bfseries 128} (2022) 032003}
  [\href{https://arxiv.org/abs/2012.12901}{{\ttfamily 2012.12901}}].

\bibitem{Aronsson:2023rli}
J.~Aronsson, D.I.~M\"uller and D.~Schuh, \emph{{Geometrical aspects of lattice
  gauge equivariant convolutional neural networks}},
  \href{https://arxiv.org/abs/2303.11448}{{\ttfamily 2303.11448}}.

\bibitem{Abbott:2022zhs}
R.~Abbott, M.S.~Albergo, D.~Boyda, K.~Cranmer, D.C.~Hackett, G.~Kanwar et~al.,
  \emph{Gauge-equivariant flow models for sampling in lattice field theories
  with pseudofermions},
  \href{https://doi.org/10.1103/PhysRevD.106.074506}{\emph{Phys. Rev. D}
  {\bfseries 106} (2022) 074506}
  [\href{https://arxiv.org/abs/2207.08945}{{\ttfamily 2207.08945}}].

\bibitem{Zhou:2023pti}
K.~Zhou, L.~Wang, L.-G.~Pang and S.~Shi, \emph{{Exploring QCD matter in extreme
  conditions with Machine Learning}},
  \href{https://arxiv.org/abs/2303.15136}{{\ttfamily 2303.15136}}.

\bibitem{He:2023zin}
W.-B.~He, Y.-G.~Ma, L.-G.~Pang, H.-C.~Song and K.~Zhou, \emph{{High-energy
  nuclear physics meets machine learning}},
  \href{https://doi.org/10.1007/s41365-023-01233-z}{\emph{Nucl. Sci. Tech.}
  {\bfseries 34} (2023) 88} [\href{https://arxiv.org/abs/2303.06752}{{\ttfamily
  2303.06752}}].

\bibitem{yang:2022diffusion}
L.~Yang, Z.~Zhang, S.~Hong, R.~Xu, Y.~Zhao, Y.~Shao et~al., \emph{Diffusion
  {{Models}}: {{A Comprehensive Survey}} of {{Methods}} and {{Applications}}},
  Sept., 2022.
\newblock 10.48550/arXiv.2209.00796.

\bibitem{Croitoru:2023dm}
F.-A.~Croitoru, V.~Hondru, R.T.~Ionescu and M.~Shah, \emph{Diffusion models in
  vision: A survey},
  \href{https://doi.org/10.1109/TPAMI.2023.3261988}{\emph{IEEE Transactions on
  Pattern Analysis and Machine Intelligence} (2023) 1}.

\bibitem{2022arXiv220406125R}
A.~{Ramesh}, P.~{Dhariwal}, A.~{Nichol}, C.~{Chu} and M.~{Chen},
  \emph{{Hierarchical Text-Conditional Image Generation with CLIP Latents}},
  {\emph{arXiv e-prints} (2022) }
  [\href{https://arxiv.org/abs/2204.06125}{{\ttfamily 2204.06125}}].

\bibitem{Rombach_2022_CVPR}
R.~Rombach, A.~Blattmann, D.~Lorenz, P.~Esser and B.~Ommer,
  \emph{High-resolution image synthesis with latent diffusion models},  in
  \emph{Proceedings of the IEEE/CVF Conference on Computer Vision and Pattern
  Recognition (CVPR)}, pp.~10684--10695, June, 2022.

\bibitem{Mikuni:2022xry}
V.~Mikuni and B.~Nachman, \emph{{Score-based generative models for calorimeter
  shower simulation}},
  \href{https://doi.org/10.1103/PhysRevD.106.092009}{\emph{Phys. Rev. D}
  {\bfseries 106} (2022) 092009}
  [\href{https://arxiv.org/abs/2206.11898}{{\ttfamily 2206.11898}}].

\bibitem{Mikuni:2023dvk}
V.~Mikuni, B.~Nachman and M.~Pettee, \emph{{Fast point cloud generation with
  diffusion models in high energy physics}},
  \href{https://doi.org/10.1103/PhysRevD.108.036025}{\emph{Phys. Rev. D}
  {\bfseries 108} (2023) 036025}
  [\href{https://arxiv.org/abs/2304.01266}{{\ttfamily 2304.01266}}].

\bibitem{Parisi:1980ys}
G.~Parisi and Y.S.~Wu, \emph{{Perturbation theory without gauge fixing}},
  {\emph{Sci. China, A} {\bfseries 24} (1980) 483}.

\bibitem{Damgaard:1987rr}
P.H.~Damgaard and H.~H{\"u}ffel, \emph{Stochastic quantization},
  \href{https://doi.org/10.1016/0370-1573(87)90144-X}{\emph{Phys. Rept.}
  {\bfseries 152} (1987) 227}.

\bibitem{Namiki:1993fd}
M.~Namiki, \emph{{Basic ideas of stochastic quantization}},
  \href{https://doi.org/10.1143/PTPS.111.1}{\emph{Prog. Theor. Phys. Suppl.}
  {\bfseries 111} (1993) 1}.

\bibitem{Parisi:1983mgm}
G.~Parisi, \emph{On complex probabilities},
  \href{https://doi.org/10.1016/0370-2693(83)90525-7}{\emph{Physics Letters B}
  {\bfseries 131} (1983) 393}.

\bibitem{Berges:2006xc}
J.~Berges, S.~Borsanyi, D.~Sexty and I.O.~Stamatescu, \emph{{Lattice
  simulations of real-time quantum fields}},
  \href{https://doi.org/10.1103/PhysRevD.75.045007}{\emph{Phys. Rev. D}
  {\bfseries 75} (2007) 045007}
  [\href{https://arxiv.org/abs/hep-lat/0609058}{{\ttfamily hep-lat/0609058}}].

\bibitem{Aarts:2008rr}
G.~Aarts and I.-O.~Stamatescu, \emph{{Stochastic quantization at finite
  chemical potential}},
  \href{https://doi.org/10.1088/1126-6708/2008/09/018}{\emph{JHEP} {\bfseries
  09} (2008) 018} [\href{https://arxiv.org/abs/0807.1597}{{\ttfamily
  0807.1597}}].

\bibitem{Aarts:2008wh}
G.~Aarts, \emph{{Can stochastic quantization evade the sign problem? The
  relativistic Bose gas at finite chemical potential}},
  \href{https://doi.org/10.1103/PhysRevLett.102.131601}{\emph{Phys. Rev. Lett.}
  {\bfseries 102} (2009) 131601}
  [\href{https://arxiv.org/abs/0810.2089}{{\ttfamily 0810.2089}}].

\bibitem{Seiler:2012wz}
E.~Seiler, D.~Sexty and I.-O.~Stamatescu, \emph{{Gauge cooling in complex
  Langevin for QCD with heavy quarks}},
  \href{https://doi.org/10.1016/j.physletb.2013.04.062}{\emph{Phys. Lett. B}
  {\bfseries 723} (2013) 213}
  [\href{https://arxiv.org/abs/1211.3709}{{\ttfamily 1211.3709}}].

\bibitem{Sexty:2013ica}
D.~Sexty, \emph{{Simulating full QCD at nonzero density using the complex
  Langevin equation}},
  \href{https://doi.org/10.1016/j.physletb.2014.01.019}{\emph{Phys. Lett. B}
  {\bfseries 729} (2014) 108}
  [\href{https://arxiv.org/abs/1307.7748}{{\ttfamily 1307.7748}}].

\bibitem{Aarts:2015tyj}
G.~Aarts, \emph{{Introductory lectures on lattice QCD at nonzero baryon
  number}}, \href{https://doi.org/10.1088/1742-6596/706/2/022004}{\emph{J.
  Phys. Conf. Ser.} {\bfseries 706} (2016) 022004}
  [\href{https://arxiv.org/abs/1512.05145}{{\ttfamily 1512.05145}}].

\bibitem{Attanasio:2020spv}
F.~Attanasio, B.~J\"ager and F.P.G.~Ziegler, \emph{{Complex Langevin
  simulations and the QCD phase diagram: Recent developments}},
  \href{https://doi.org/10.1140/epja/s10050-020-00256-z}{\emph{Eur. Phys. J. A}
  {\bfseries 56} (2020) 251}
  [\href{https://arxiv.org/abs/2006.00476}{{\ttfamily 2006.00476}}].

\bibitem{Berger:2019odf}
C.E.~Berger, L.~Rammelm\"uller, A.C.~Loheac, F.~Ehmann, J.~Braun and J.E.~Drut,
  \emph{{Complex Langevin and other approaches to the sign problem in quantum
  many-body physics}},
  \href{https://doi.org/10.1016/j.physrep.2020.09.002}{\emph{Phys. Rept.}
  {\bfseries 892} (2021) 1} [\href{https://arxiv.org/abs/1907.10183}{{\ttfamily
  1907.10183}}].

\bibitem{Nagata:2021ugx}
K.~Nagata, \emph{{Finite-density lattice QCD and sign problem: Current status
  and open problems}},
  \href{https://doi.org/10.1016/j.ppnp.2022.103991}{\emph{Prog. Part. Nucl.
  Phys.} {\bfseries 127} (2022) 103991}
  [\href{https://arxiv.org/abs/2108.12423}{{\ttfamily 2108.12423}}].

\bibitem{Aarts:2009uq}
G.~Aarts, E.~Seiler and I.-O.~Stamatescu, \emph{Complex {{Langevin}} method:
  {{When}} can it be trusted?},
  \href{https://doi.org/10.1103/PhysRevD.81.054508}{\emph{Phys. Rev. D}
  {\bfseries 81} (2010) 054508}.

\bibitem{Aarts:2011ax}
G.~Aarts, F.A.~James, E.~Seiler and I.-O.~Stamatescu, \emph{{Complex Langevin:
  Etiology and Diagnostics of its Main Problem}},
  \href{https://doi.org/10.1140/epjc/s10052-011-1756-5}{\emph{Eur. Phys. J. C}
  {\bfseries 71} (2011) 1756}
  [\href{https://arxiv.org/abs/1101.3270}{{\ttfamily 1101.3270}}].

\bibitem{Nagata:2016vkn}
K.~Nagata, J.~Nishimura and S.~Shimasaki, \emph{Argument for justification of
  the complex {{Langevin}} method and the condition for correct convergence},
  \href{https://doi.org/10.1103/PhysRevD.94.114515}{\emph{Phys. Rev. D}
  {\bfseries 94} (2016) 114515}.

\bibitem{Aarts:2017vrv}
G.~Aarts, E.~Seiler, D.~Sexty and I.-O.~Stamatescu, \emph{{Complex Langevin
  dynamics and zeroes of the fermion determinant}},
  \href{https://doi.org/10.1007/JHEP05(2017)044}{\emph{JHEP} {\bfseries 05}
  (2017) 044} [\href{https://arxiv.org/abs/1701.02322}{{\ttfamily
  1701.02322}}].

\bibitem{Scherzer:2018hid}
M.~Scherzer, E.~Seiler, D.~Sexty and I.-O.~Stamatescu, \emph{{Complex Langevin
  and boundary terms}},
  \href{https://doi.org/10.1103/PhysRevD.99.014512}{\emph{Phys. Rev. D}
  {\bfseries 99} (2019) 014512}
  [\href{https://arxiv.org/abs/1808.05187}{{\ttfamily 1808.05187}}].

\bibitem{WesthHansen:2022iqd}
M.~Westh~Hansen and D.~Sexty, \emph{{Complex Langevin boundary terms in full
  QCD}}, \href{https://doi.org/10.22323/1.430.0163}{\emph{PoS} {\bfseries
  LATTICE2022} (2023) 163} [\href{https://arxiv.org/abs/2212.12029}{{\ttfamily
  2212.12029}}].

\bibitem{Alvestad:2021hsi}
D.~Alvestad, R.~Larsen and A.~Rothkopf, \emph{{Stable solvers for real-time
  Complex Langevin}},
  \href{https://doi.org/10.1007/JHEP08(2021)138}{\emph{JHEP} {\bfseries 08}
  (2021) 138} [\href{https://arxiv.org/abs/2105.02735}{{\ttfamily
  2105.02735}}].

\bibitem{Alvestad:2022abf}
D.~Alvestad, R.~Larsen and A.~Rothkopf, \emph{{Towards learning optimized
  kernels for complex Langevin}},
  \href{https://doi.org/10.1007/JHEP04(2023)057}{\emph{JHEP} {\bfseries 04}
  (2023) 057} [\href{https://arxiv.org/abs/2211.15625}{{\ttfamily
  2211.15625}}].

\bibitem{Lampl:2023xpb}
N.M.~Lampl and D.~Sexty, \emph{{Real time evolution of scalar fields with
  kernelled Complex Langevin equation}},
  \href{https://arxiv.org/abs/2309.06103}{{\ttfamily 2309.06103}}.

\bibitem{Callaway:1982eb}
D.J.E.~Callaway and A.~Rahman, \emph{{The Microcanonical Ensemble: A New
  Formulation of Lattice Gauge Theory}},
  \href{https://doi.org/10.1103/PhysRevLett.49.613}{\emph{Phys. Rev. Lett.}
  {\bfseries 49} (1982) 613}.

\bibitem{risken1996fokker}
H.~Risken, \emph{The Fokker-Planck equation: Methods of solution and
  application}, Springer (1996).

\bibitem{Namiki:1992wf}
M.~Namiki, I.~Ohba, K.~Okano, Y.~Yamanaka, A.K.~Kapoor, H.~Nakazato et~al.,
  \emph{{Stochastic quantization}}, vol.~9, Springer Berlin Heidelberg (1992),
  \href{https://doi.org/10.1007/978-3-540-47217-9}{10.1007/978-3-540-47217-9}.

\bibitem{Batrouni:1985jn}
G.G.~Batrouni, G.R.~Katz, A.S.~Kronfeld, G.P.~Lepage, B.~Svetitsky and
  K.G.~Wilson, \emph{{Langevin Simulations of Lattice Field Theories}},
  \href{https://doi.org/10.1103/PhysRevD.32.2736}{\emph{Phys. Rev. D}
  {\bfseries 32} (1985) 2736}.

\bibitem{sohl-dickstein:2015deep}
J.~{Sohl-Dickstein}, E.A.~Weiss, N.~Maheswaranathan and S.~Ganguli, \emph{Deep
  unsupervised learning using nonequilibrium thermodynamics},  in \emph{Proc.
  32nd {{Int}}. {{Conf}}. {{Int}}. {{Conf}}. {{Mach}}. {{Learn}}. - {{Vol}}.
  37}, {{ICML}}'15, ({Lille, France}), pp.~2256--2265, {JMLR.org}, July, 2015.

\bibitem{NEURIPS2019_3001ef25}
Y.~Song and S.~Ermon, \emph{Generative modeling by estimating gradients of the
  data distribution},  in \emph{Adv. {{Neural Inf}}. {{Process}}. {{Syst}}.},
  H.~Wallach, H.~Larochelle, A.~Beygelzimer, F.~{dAlch{\'e}-Buc}, E.~Fox and
  R.~Garnett, eds., vol.~32, {Curran Associates, Inc.}, 2019.

\bibitem{song2021scorebased}
Y.~Song, J.~{Sohl-Dickstein}, D.P.~Kingma, A.~Kumar, S.~Ermon and B.~Poole,
  \emph{Score-based generative modeling through stochastic differential
  equations},  in \emph{Int. {{Conf}}. {{Learn}}. {{Represent}}.}, 2021.

\bibitem{ho:2020denoising}
J.~Ho, A.~Jain and P.~Abbeel, \emph{Denoising diffusion probabilistic models},
  in \emph{Proc. 34th {{Int}}. {{Conf}}. {{Neural Inf}}. {{Process}}.
  {{Syst}}.}, {{NIPS}}'20, ({Red Hook, NY, USA}), pp.~6840--6851, {Curran
  Associates Inc.}, Dec., 2020.

\bibitem{anderson:1982reversetime}
B.D.O.~Anderson, \emph{Reverse-time diffusion equation models},
  \href{https://doi.org/10.1016/0304-4149(82)90051-5}{\emph{Stochastic
  Processes and their Applications} {\bfseries 12} (1982) 313}.

\bibitem{Aapo:2005scb}
A.~Hyv{{\"a}}rinen, \emph{Estimation of non-normalized statistical models by
  score matching}, {\emph{Journal of Machine Learning Research} {\bfseries 6}
  (2005) 695}.

\bibitem{Vincent:2011scb}
P.~Vincent, \emph{A connection between score matching and denoising
  autoencoders}, \href{https://doi.org/10.1162/NECO_a_00142}{\emph{Neural
  Computation} {\bfseries 23} (2011) 1661}.

\bibitem{welling:2011bayesian}
M.~Welling and Y.W.~Teh, \emph{Bayesian learning via stochastic gradient
  langevin dynamics},  in \emph{Proc. 28th {{Int}}. {{Conf}}. {{Int}}.
  {{Conf}}. {{Mach}}. {{Learn}}.}, {{ICML}}'11, ({Madison, WI, USA}),
  pp.~681--688, {Omnipress}, June, 2011.

\bibitem{Maoutsa:2020ode}
D.~Maoutsa, S.~Reich and M.~Opper, \emph{Interacting particle solutions of
  fokker–planck equations through gradient–log–density estimation},
  {\emph{Entropy} {\bfseries 22} (2020) }.

\bibitem{grathwohl2018scalable}
W.~Grathwohl, R.T.Q.~Chen, J.~Bettencourt and D.~Duvenaud, \emph{Scalable
  reversible generative models with free-form continuous dynamics},  in
  \emph{International Conference on Learning Representations}, 2019.

\bibitem{Luscher:2009eq}
M.~Luscher, \emph{{Trivializing maps, the Wilson flow and the HMC algorithm}},
  \href{https://doi.org/10.1007/s00220-009-0953-7}{\emph{Commun. Math. Phys.}
  {\bfseries 293} (2010) 899}
  [\href{https://arxiv.org/abs/0907.5491}{{\ttfamily 0907.5491}}].

\bibitem{Bern:1986xc}
Z.~Bern, M.B.~Halpern and L.~Sadun, \emph{{Continuum Regularization of Quantum
  Field Theory. 4. Langevin Renormalization}},
  \href{https://doi.org/10.1007/BF01408455}{\emph{Z. Phys. C} {\bfseries 35}
  (1987) 255}.

\bibitem{Pawlowski:2017rhn}
J.M.~Pawlowski, I.-O.~Stamatescu and F.P.G.~Ziegler, \emph{{Cooling Stochastic
  Quantization with colored noise}},
  \href{https://doi.org/10.1103/PhysRevD.96.114505}{\emph{Phys. Rev. D}
  {\bfseries 96} (2017) 114505}
  [\href{https://arxiv.org/abs/1705.06231}{{\ttfamily 1705.06231}}].

\bibitem{Smit:2002ug}
J.~Smit, \emph{{Introduction to quantum fields on a lattice: A robust mate}},
  vol.~15, Cambridge University Press (1, 2011).

\bibitem{Akiyama:2021zhf}
S.~Akiyama, Y.~Kuramashi and Y.~Yoshimura, \emph{{Phase transition of
  four-dimensional lattice ${\phi}^4$ theory with tensor renormalization
  group}}, \href{https://doi.org/10.1103/PhysRevD.104.034507}{\emph{Phys. Rev.
  D} {\bfseries 104} (2021) 034507}
  [\href{https://arxiv.org/abs/2101.06953}{{\ttfamily 2101.06953}}].

\bibitem{neal2011mcmc}
R.M.~Neal et~al., \emph{Mcmc using hamiltonian dynamics}, {\emph{Handbook of
  markov chain monte carlo} {\bfseries 2} (2011) 2}.

\bibitem{Ronneberger:2015unet}
O.~Ronneberger, P.~Fischer and T.~Brox, \emph{U-net: Convolutional networks for
  biomedical image segmentation},  in \emph{Medical Image Computing and
  Computer-Assisted Intervention -- MICCAI 2015}, N.~Navab, J.~Hornegger,
  W.M.~Wells and A.F.~Frangi, eds., (Cham), pp.~234--241, Springer
  International Publishing, 2015.

\bibitem{Binder:1981zz}
K.~Binder, \emph{{Critical Properties from Monte Carlo Coarse Graining and
  Renormalization}},
  \href{https://doi.org/10.1103/PhysRevLett.47.693}{\emph{Phys. Rev. Lett.}
  {\bfseries 47} (1981) 693}.

\bibitem{Bachtis:2020ajb}
D.~Bachtis, G.~Aarts and B.~Lucini, \emph{{Mapping distinct phase transitions
  to a neural network}},
  \href{https://doi.org/10.1103/PhysRevE.102.053306}{\emph{Phys. Rev. E}
  {\bfseries 102} (2020) 053306}
  [\href{https://arxiv.org/abs/2007.00355}{{\ttfamily 2007.00355}}].

\bibitem{Schaefer:2010hu}
{\scshape ALPHA} collaboration, \emph{{Critical slowing down and error analysis
  in lattice QCD simulations}},
  \href{https://doi.org/10.1016/j.nuclphysb.2010.11.020}{\emph{Nucl. Phys. B}
  {\bfseries 845} (2011) 93} [\href{https://arxiv.org/abs/1009.5228}{{\ttfamily
  1009.5228}}].

\bibitem{Wolff:2003sm}
{\scshape ALPHA} collaboration, \emph{{Monte Carlo errors with less errors}},
  \href{https://doi.org/10.1016/S0010-4655(03)00467-3}{\emph{Comput. Phys.
  Commun.} {\bfseries 156} (2004) 143}
  [\href{https://arxiv.org/abs/hep-lat/0306017}{{\ttfamily hep-lat/0306017}}].

\bibitem{Joswig:2022qfe}
F.~Joswig, S.~Kuberski, J.T.~Kuhlmann and J.~Neuendorf, \emph{{pyerrors: A
  python framework for error analysis of Monte Carlo data}},
  \href{https://doi.org/10.1016/j.cpc.2023.108750}{\emph{Comput. Phys. Commun.}
  {\bfseries 288} (2023) 108750}
  [\href{https://arxiv.org/abs/2209.14371}{{\ttfamily 2209.14371}}].

\bibitem{Robnik:2023pgt}
J.~Robnik and U.~Seljak, \emph{{Microcanonical Langevin Monte Carlo}},
  \href{https://arxiv.org/abs/2303.18221}{{\ttfamily 2303.18221}}.

\bibitem{Cotler:2023lem}
J.~Cotler and S.~Rezchikov, \emph{{Renormalizing Diffusion Models}},
  \href{https://arxiv.org/abs/2308.12355}{{\ttfamily 2308.12355}}.

\end{thebibliography}\endgroup

\end{document}